\documentclass[12pt]{article}
\usepackage[dvips]{graphicx}
\usepackage{amsfonts}
\usepackage{amsmath}
\usepackage{amssymb}

\usepackage[T1]{fontenc}
\usepackage[latin1]{inputenc}
\usepackage{graphicx}
\usepackage{epsfig}

\newtheorem{prop}{Proposition}[section]

\newtheorem{Remark}{Remark}[section]

\begin{document}
\title{Interference effects at electron tunneling}

\author{N. Angelescu\footnote{Institute for Physics and Nuclear Engineering "H. Hulubei", P.O.Box MG-6, Bucharest, Romania; e-mail: {\tt nangel@theory.nipne.ro}}\, , M. Bundaru\footnote{Institute for Physics and Nuclear Engineering "H. Hulubei", P.O.Box MG-6, Bucharest, Romania; e-mail: {\tt bundaru@theory.nipne.ro}}\, , R. Bundaru\thanks{Institute for Space Sciences, P.O. Box MG-23,
Bucharest, Romania, e-mail: {\tt bundaru@spacescience.ro}}\, and I. Popescu \thanks{Institute for Space Sciences, P.O. Box MG-23,
Bucharest, Romania}}

\date{}

\maketitle

\begin{abstract}
In this article we discuss the interference patterns which appear in
the local particle-density and particle-current distributions of two
infinite 2-D lattice systems of free fermions which are allowed two
communicate via two tunneling junction points. The two fermion
reservoirs are initially in different invariant states, and the
particle-density and particle-current distributions are calculated
in the final stationary state. The dependence of the interference
pattern on the interaction details is discussed in several examples.
\end{abstract}

\section{Introduction}\label{s1}

The paper is concerned with a simple model of electron tunneling
between infinite reservoirs in different equilibrium states. We work
in a tight binding approximation, i.e. the electrons live on
lattices and the one-particle kinetic energy is given by appropriate
lattice Laplace operators. Thereby, the interactions are neglected,
so that the Hamilton operators of the isolated reservoirs are
bilinear in the electron creation/annihilation operators. At time
$t=0$ a coupling between the reservoirs is switched on, likewise
bilinear, which allows direct hopping of the particles between
certain finite subsets of sites in different reservoirs.
Asymptotically in time, the system approaches a stationary state, in
which permanent currents of particles and energy are present. For
such models, the approach to stationarity and the structure of the
stationary state is well understood at the mathematical level of
rigor \cite{AJPP}, \cite{abb}. The stationary state is a quasi-free
state \cite{BR}, which is completely characterized in terms of the
spectral properties of the one-particle Hamilton operators of the
free and coupled reservoirs. Explicit formulas for the permanent
particle and energy currents between systems are known (the
Landauer-B\"{u}ttiker formulas). However, the state provides more
than such global information about the system: it allows obtaining
the expectation values of all the local observables.

Our aim here is to perform such a detailed study of the stationary
state, by calculating the expectations of various local observables
(namely, the local particle density and the particle currents) for a
particular geometric arrangement and tunneling Hamiltonian. More
precisely, we take the reservoirs to be free fermion systems living
on regular two-dimensional lattices, with direct tunneling between
(one- and/or two-site) finite subsets of the latter. Such choice of
the reservoirs is by no means unrealistic, e.g. electrons in the
surface states of the close-packed metal surfaces are
two-dimensional, nearly free, electron gases. Moreover, the local
density of electrons is experimentally accessible by scanning
tunneling microscopy, by a technique elaborated in the nineties by
Crommie et al \cite{STM1}. Standing density waves were observed on
the Cu(111) surface and their origin explained by the scattering on
point impurities located on the surface \cite{STM1},\cite{STM2}.
Such measurements have since been performed on different materials
and defect configurations (see \cite{nano} for a recent experiment),
and various applications proposed (e.g. \cite{focus}). Though our
calculations refer to a different physical situation, we hope that
they are not only of theoretical interest and that the interference
effects exhibited by us can be experimentally observed.

The paper is organized as follows. The precise description of the
model and its general properties are done in Section \ref{s2}. In
order to establish notation and for reader's convenience, the
approach to, and structure of, the stationary state are presented in
Subsection \ref{s2.1}, following \cite{AJPP}, \cite{abb}. The main
ingredient is the perturbation analysis of the coupled Hamiltonian.
As we want to demonstrate the appearance of quantum interference
effects in the stationary state, we derive the formulas for the
particle density profile (Subsection \ref{s2.2}) and the local
particle currents (Subsection \ref{s2.3}) in the stationary state.
Section \ref{s3} is devoted to numerical illustrations. Here, we
specify the reservoirs to be two-dimensional free fermion lattice
gases, with direct tunneling between (one- and/or two-site) finite
subsets of the two square lattices. Subsection \ref{s3.1} contains
the numerical study of the limit values on the real axis of the
Green's function of the two-dimensional lattice Laplacian, which
enter the expression of the stationary state. Subsections
\ref{s3.2}, \ref{s3.3} are devoted to numerical calculations of the
charge- and of the current-density profiles, respectively, for the
particular geometry described above. Classically, the total current
should be a sum of the currents established through each individual
channel when the other are closed. The computations show that this
is no longer true in the quantum case. Moreover, the calculated
charge-density profiles in the reservoirs exhibit interference
fringes, depending on the tunneling constants for every pair of
coupled sites. Section \ref{s4} is devoted to a discussion of the
results.

\section{The stationary state}\label{s2}
\setcounter{equation}{0}

\subsection{The model and the approach to stationarity}\label{s2.1}
We consider two lattice free electron gases on
$\mathbb{L}_i=\mathbb{Z}^{d_i}$,
 $i=1,2$ (where $\mathbb{Z}$ denotes the integers), coupled via a tunneling
junction of finite support. A one-electron wave function is a vector
$f\in {\cal H}=l^2(\mathbb{L}_1\cup
\mathbb{L}_2)=l^2(\mathbb{L}_1)\oplus l^2(\mathbb{L}_2)$, i.e. $f=\{
f_x;\; x\in \mathbb{L}_1\cup \mathbb{L}_2\}$, such that $\| f\|
^2:=\sum\limits_{x\in \mathbb{L}_1\cup \mathbb{L}_2}|f_x|^2<\infty$.
Let $a^*_x,\; a_x$ be the operator of creation, respectively
annihilation, of an electron at the site $x$ acting in the
anti-symmetric Fock space $\cal F$ over $\cal H$. Also, for $f\in
\cal H$, we denote $a^*(f)=\sum _xf_xa^*_x$ and $a(f)=\sum _x
\overline{f_x}a_x$. For a one-particle operator $c$, given by the
matrix $\left( c_{x,y}\right)$ ($x,y\in \mathbb{L}_1\cup
\mathbb{L}_2$), its second-quantization, acting in $\cal F$, is
defined as $${\rm d}\Gamma (c)=\sum\limits_{x,y\in \mathbb{L}_1\cup
\mathbb{L}_2}c_{x,y}a^*_xa_y.$$

\bigskip
Putting the electron mass equal to 1, the one-particle Hamiltonian
$h^0$ corresponding to uncoupled reservoirs, which represents the
kinetic energy operator, is given by
\begin{equation}\label{2.1}
    (h^0f)_x=\sum\limits_{i=1}^2(h^0_if)_x=-\frac{1}{2}\sum\limits_{i=1}^2\chi _{\mathbb{L}_i}(x)\sum\limits_{y\in \mathbb{L}_i,\,
    |y-x|=1}(f_y-f_x),
\end{equation}
where $\chi _A(x)$ denotes the characteristic function of the set
$A$. The evolution in time (in the Heisenberg picture) of $a^\sharp
(f)$ writes as $$\alpha ^0_t(a^\sharp (f)):={\rm e}^{{\rm i}t{\rm
d}\Gamma (h^0)}a^\sharp (f){\rm e}^{-{\rm i}t{\rm d}\Gamma
(h^0)}=a^\sharp ({\rm e}^{ih^0t}f).$$

The generalized eigenfunctions $\phi (k)$ of $h^0$ are plane waves
in each of the two reservoirs, corresponding to given momentum $k\in
\mathbb{T}_1\cup \mathbb{T}_2$, where $\mathbb{T}_i=(-\pi ,\pi
]^{d_i}$:
\begin{equation}\label{2.2}
    {\rm For}\;\; k\in \mathbb{T}_i,\;\;\; \phi (k)_x=\chi _{\mathbb{L}_i}(x)
                         (2\pi )^{-d_i/2}\exp{(ikx)},
    \;\;\; (i=1,2).
\end{equation}
One has $h^0\phi (k)=\omega (k)\phi (k)$, where the dispersion law
is
\begin{equation}\label{2.3}
    \omega (k)
    =\sum\limits_{i=1}^2\chi _{\mathbb{T}_i}(k)\sum\limits_{\alpha=1}^{d_i}2\sin ^2{k_\alpha
    /2}.
\end{equation}

At time $t\leq 0$, the two (uncoupled) reservoirs are supposed to be
in different equilibria, defined by inverse temperatures $\beta _i$
and chemical potentials $\mu _i$, $i=1,2$. Let $<\cdot
>_0$ denote this constrained equilibrium state. The density operator of the state $<\cdot
>_0$ is
\begin{equation}\label{2.4}
    (\rho ^0)_{x,y}:=<a^*_xa_y>_0=\sum\limits_{i=1}^2\int\limits_{k\in \mathbb{T}_i}{\rm d}k\phi
  (k)_x\overline{\phi (k)_y}  f_{\beta _i,\mu _i}(\omega (k))=\sum\limits_{i=1}^2\int\limits_{k\in \mathbb{T}_i}{\rm d}k E(k)_{x,y}f_{\beta _i,\mu _i}(\omega
  (k)),
\end{equation}
where $f_{\beta ,\mu}(x)=(1+\textrm{e}^{\beta (x-\mu )})^{-1}$ is
the Fermi function, and where we defined the matrix $E(k)$ by
\begin{equation}\label{2.5}
    E(k)_{x,y}=\phi (k)_x\overline{\phi (k)_y},\;\; x,y\in \mathbb{L}_1\cup \mathbb{L}_2.
\end{equation}

The higher order correlations in the state $<\cdot >_0$ are
calculated in terms of these as:
\begin{equation}\label{2.6}
    <a^*(f_1)...a^*(f_m)a(g_1)...a(g_n)>_0=\delta_{m,n}\det{M},
\;\;\; M_{i,j}=<a^*(f_i)a(g_j)>_0=(g_j,\rho ^0f_i)_{\cal H}.
\end{equation}
A state with the latter property is called \emph{quasi-free and
gauge-invariant}.

\bigskip
Starting from $t=0$, a tunneling junction is opened, connecting the
finite subsets $S_i\in \mathbb{L}_i$, $i=1,2$. The future time
evolution is controlled by the perturbed one-particle Hamiltonian
$h=h^0+v$, where $v=\tau +\tau ^*$ with $\tau
:l^2(\mathbb{L}_2)\rightarrow l^2(\mathbb{L}_1)$ describing the
tunneling of particles from the second to the first reservoir, i.e.
\begin{equation}\label{2.7}
    (\tau f)_x=\chi _{S_1}(x)\sum\limits_{y\in S_2}t_{x,y}f(y),
\end{equation}
and $\tau ^*:l^2(\mathbb{L}_1)\rightarrow l^2(\mathbb{L}_2)$ the
tunneling in the opposite direction
$(t^*)_{x,y}=\overline{t_{y,x}}$. The Heisenberg evolution of
$a^\sharp (f)$ for $t>0$ writes \[ \alpha _t(a^\sharp (f)):={\rm
e}^{{\rm i}t{\rm d}\Gamma (h^0+v)}a^\sharp (f){\rm e}^{-{\rm i}t{\rm
d}\Gamma (h^0+v)}=a^\sharp ({\rm e}^{i(h^0+v)t}f).\]

The initial state $<\cdot >_0$ is no longer invariant under the new
evolution. At time $t>0$, the state will be the gauge-invariant,
quasi-free state $<\cdot >_t$ of density operator
\begin{equation}\label{2.8}
    \rho ^t=[{\rm e}^{-{\rm i}th^0}{\rm e}^{{\rm i}th}]^*\rho ^0
    {\rm e}^{-{\rm i}th^0}{\rm e}^{{\rm i}th};
\end{equation}
here, the factors ${\rm e}^{{\rm i}th^0}$ are added for convenience
(they commute with $\rho ^0$, hence do not contribute), as they show
that, in far future, the state is given by the M\"{o}ller operators
in the scattering theory for the perturbed evolution. More
precisely, we shall use the following result, which is an easy
consequence of the general results of \cite{AJPP}, \cite{abb}:
\begin{prop}
In the framework defined above,
\begin{enumerate}
  \item the spectrum of $h$ consists of
absolutely continuous spectrum, equal to the spectrum of $h^0$,
$\sigma _{\rm ac}(h)=\sigma (h^0)=\cup _{i=1,2}[0,2d_i]$ and of a
finite set of eigenvalues, $\sigma _{\rm p}(h)$, outside $\sigma
(h^0)$; thereby, the wave operators
\begin{equation}\label{2.9}
   W_\pm := {\rm s}-\lim\limits_{t\to \pm \infty}{\rm e}^{{\rm i}th}{\rm e}^{-{\rm i}th^0}
\end{equation}
exist and establish the unitary equivalence of $h^0$ with the
absolutely continuous part $h_{ac}$ of $h$: $h_{ac}W_\pm =W_\pm
h^0$;
  \item the states $<\cdot >_t$ approach in mean, as $t\to \infty$, a
  stationary state, i.e. the following limits exist:
\begin{equation}\label{2.10}
    \lim\limits _{T\to +\infty }(1/T)\int _0^T <A>_t{\rm
    d}t=:<A>_{\rm stat},
\end{equation}
    for all polynomials $A$ in $a^\sharp (f)$;
    \item
  $<\cdot >_{\rm stat}$ is the gauge-invariant,
  quasi-free state of density operator
  \begin{equation}\label{2.11}
    \rho _+=W_-\rho ^0W_-^*+\sum\limits _{\lambda  \in \sigma _{\rm p}(h)}P_\lambda \rho ^0P_\lambda ,
  \end{equation}
    where $P_\lambda$ is the projection onto the eigenspace of $h$
    corresponding to the eigenvalue $\lambda$.
\end{enumerate}
\end{prop}

\bigskip
The rest of this section is devoted to deriving computable
expressions for the eigenvectors and of the M\"{o}ller operators.

We start with writing the resolvent operators $r^0(z)=(h^0-z)^{-1}$
and  $r(z)=(h-z)^{-1}$. Clearly, $r^0(z)$ is expressed in terms of
the generalized eigenfunctions (\ref{2.2}) of $h^0$ as:
\begin{equation}\label{2.12}
    r^0(z)_{x,y}=\int\limits_{\mathbb{T}_1\cup\mathbb{T}_2}\frac{E(k)_{x,y}}{\omega
    (k)-z}{\rm d}k=\int \frac{P(e)_{x,y}}{e-z}{\rm d}e,
\end{equation}
where $E(k)$ is defined in Eq.(\ref{2.5}) and  $P(e)$ is the
integral of $E(k)$ over the energy shell $\Sigma _1(e)\cup \Sigma
_2(e)$, with $$\Sigma _i(e)=\{ k\in \mathbb{T}_i;\; \omega
(k)=e\}.$$ Both $E(k)$ and $P(e)$ have block diagonal structure with
respect to the reservoirs. In particular, $P(e)=P_1(e)\oplus P_2(e)$
where, denoting ${\rm d}\sigma _i(k)$ the element of surface area on
$\Sigma _i(e)$, we have, explicitly,
\begin{equation}\label{2.13}
P(e)_{x,y}=\sum\limits_{i=1}^2\chi
_{\mathbb{L}_i}(x-y)P_i(e)_{x,y}=\sum\limits_{i=1}^2\chi
_{\mathbb{L}_i}(x-y)\int\limits_{\Sigma _i(e)}\frac{{\rm d}\sigma
_i(k)}{|\nabla \omega (k)|}E(k)_{x,y}.
\end{equation}
Thereby, $ r^0(z)_{x,y}=\sum\limits_{i=1}^2\chi
_{\mathbb{L}_i}(x-y)g_i(z,y-x)$, where $g_i(z,\cdot):\mathbb{L}_i\to
\mathbb{C}$ is defined as
\begin{equation}\label{2.14}
g_i(z,x)=\int\limits_0^{2d_i} \frac{P_i(e)_{0,x}}{e-z}{\rm d}e.
\end{equation}
\medskip
To calculate $r(z)$, one has to solve for $f$ the equation
$(h^0-z+v)f=g$. If $z\in \mathbb{C}\setminus \sigma (h^0)$, this is
equivalent to $f+r^0(z)vf=r^0(z)g$. Applying $v$ to the latter
equation, one gets $(v+vr^0(z)v)f=vr^0(z)g$. Let $\Pi$ be the
projection on the range of $v$: $\Pi {\cal H}=(\ker{v})^\bot$. There
are two possibilities:
\begin{itemize}
  \item {\em The restriction of $v+vr^0(z)v$ to $\Pi {\cal H}$ is invertible}.
  Then, $\Pi f=(v+vr^0(z)v)^{-1}vr^0(z)g$. Once $\Pi f$ is known, the whole vector $f$ is
$f=-r^0(z)v(\Pi f)+r^0(z)g$. For such values of $z$, $h-z$ has a
bounded inverse $r(z)$:
\begin{equation}\label{2.15}
    r(z)=r^0(z)-r^0(z)Q(z)r^0(z),\;\; {\rm where}\;\;\;
    Q(z)=v(v+vr^0(z)v)^{-1}v;
\end{equation}
  \item {\em There exists a non-zero vector $\psi \in \Pi {\cal H}$, such that
  $(v+vr^0(z)v)\psi =0$}. Such values of $z$ are eigenvalues of $h$ with
  eigenvector $f=-r^0(z)v\psi$. Conversely, if $f$ is an eigenvector
  of $h$ for the eigenvalue $z$, then $\Pi f\neq 0$ and $f+r^0(z)vf=0$. Applying
  $v$, it follows that $(v+vr^0(z)v)f=(v+vr^0(z)v)(\Pi f)=0$.
  Therefore,  $\sigma
  _{\rm p}(h)$ equals the set of $z$ with this property and there is a one-to-one correspondence between the
  eigenspace corresponding to $z$ and $\Pi {\cal H}\cap
  \ker{(v+vr^0(z)v)}$.
\end{itemize}

We consider next the M\"{o}ller operator $W_-$ entering Eq.
(\ref{2.11}). As a prerequisite, we remark that the functions
$g_i(\cdot ;x)$ defined in Eq. (\ref{2.14}) are analytic in the
complex plane cut along the segment $[0,2d_i]$ and have finite
boundary values, i.e. the limits
\begin{equation}\label{2.16}
    \lim\limits_{\epsilon \searrow 0}g_i(e\pm {\rm i}\epsilon ,x)=:g_{i,\pm }(e;x)
\end{equation}
exist for all $e\in (0,2d_i)$ and $x\in \mathbb{L}_i$. As a
consequence, and because $v$ has finite range $\Pi {\cal H}$, the
limit values
\[ Q_\pm (e):=\lim\limits_{\epsilon \searrow 0}Q(e\pm {\rm i}\epsilon)
\]
exist. Clearly, $Q_\pm (e)_{x,y}=0$ unless $x,y\in S_1\cup S_2$, so
we shall view $Q_\pm (e)$ as a matrix indexed by $S_1\cup S_2$, and
denote $Q_\pm (e)^{(i,j)}_{x,y}$ its submatrices corresponding to
$x\in S_i$ and $y\in S_j$ ($i,j=1,2$).


Applying $W_-$, Eq. (\ref{2.9}), to a generalized eigenfunction of
$h^0$, one has:
\begin{equation}\label{2.17}
\begin{array}{ll}
[W_-\phi (k)]_x= & \lim\limits_{\epsilon \searrow
    0}\int _0^\infty \epsilon \left( \delta _x,\; {\rm e}^{-{\rm i}t(h-\omega (k)-i\epsilon )}\phi
    (k)\right) {\rm d}t \\ & =\lim\limits_{\epsilon \searrow
    0}\left( \delta _x,\; (-{\rm i}\epsilon r(\omega (k)+i\epsilon )\phi
    (k)\right) \\ & = \phi
    (k)_x-[r^0_+(\omega (k))Q_+(\omega (k))\phi
    (k)]_x \\ & \\ & =(\delta _x,[I-r^0_+(\omega (k))Q_+(\omega (k))]\phi (k)),
\end{array}
\end{equation}
where we used Eq.(\ref{2.15}) for $r(z)$ and the obvious relation
$(-{\rm i}\epsilon r^0(\omega (k')+i\epsilon )\phi (k')=\phi (k')$.
Likewise,
\begin{equation}\label{2.18}
    [W^*_-\phi (k)]_x =(\delta _x,[I-Q_-(\omega (k))r^0_-(\omega (k))]\phi (k)).
\end{equation}

\bigskip
We shall demonstrate the interference effects on two properties of
the stationary state $<>_{\rm stat}$, the permanent currents and the
particle density. In the calculations below, \emph{we make the
assumption that $t_{x,y}$ (hence, all $h_{x,y}$) are real numbers}
(physically, complex $h_{x,y}$ are related to the presence of
magnetic fields).

\subsection{The permanent currents in the stationary state}\label{s2.2}

Let $N_{\{ x\} }=a^*_xa_x={\rm d}\Gamma (\chi _x)$ be the particle
number operator at the site $x\in \mathbb{L}_1\cup \mathbb{L}_2$,
where $\chi _x$ is the projection onto the vector $\delta _x$, i.e.
its matrix is $(\chi _x)_{y,z}=\delta _{x,y}\delta _{x,z}$. If the
evolution is given by the one-particle Hamiltonian $h$, the flux of
particles from the site $x$ is defined as $$I_{\{ x\}}:=-\frac{{\rm
d}}{{\rm d}t}\alpha _t({\rm d}\Gamma (\chi _x)|_{t=0}=-{\rm i}[{\rm
d}\Gamma (h),{\rm d}\Gamma (\chi _x)]={\rm d}\Gamma ([-{\rm i}h,\chi
_x]).$$ In the stationary state, as $<\alpha _t(N_{\{ x\} })>_{\rm
stat}$ is independent of time, we have the "continuity equation",
saying that the total current flowing from the site $x$ vanishes:
\[ J_{\{ x\}}:=
<{\rm d}\Gamma ([-{\rm i}h,\chi _x])>_{\rm stat}=0.
\]
The l.h.s. of this equation equals $\sum\limits_{y\in
\mathbb{L}_1\cup \mathbb{L}_2}2\Im{<h_{y,x}a^*_ya_x>_{\rm stat}}$.
We define the current along the oriented bond $\{ x,y\}$ in the
stationary state as:
\begin{equation}\label{2.19}
    j_{x,y}=2\Im{<h_{y,x}a^*_ya_x>_{\rm
stat}}=2h_{y,x}\Im{(\delta _x,\rho _+\delta _y)}.
\end{equation}
Clearly, $j_{x,y}=-j_{y,x}$ and $j_{x,y}=0$ unless $h_{x,y}\neq 0$,
what happens, under our assumptions, only if either $h^0_{x,y}\neq
0$, implying that $x,y$ are nearest neighbors in one of the two
lattices, or $v_{x,y}\neq 0$, implying that $x\in S_1$ and $y\in
S_2$, or  $x\in S_2$ and $y\in S_1$.

Note that, in view of the fact that all $h_{y,x}$ are real, the
eigenvectors of $h$ can be chosen to have real components. It
follows that \emph{the point spectrum (i.e. the second term in Eq.
{\rm (\ref{2.11})}) does not contribute to $j_{x,y}$}, and Eq.
(\ref{2.19}) writes
\begin{equation}\label{2.20}
    j_{x,y}=2h_{x,y}\Im{(W_-^*\delta _x,\rho ^0W_-^*\delta _y)}=2h_{x,y}\int\limits_{\mathbb{T}_1\cup \mathbb{T}_2}
    \rho ^0(k)\Im{\{
    (W_-^*\delta _x,E(k)W_-^*\delta _y)\} }{\rm d}k.
\end{equation}

Let $\Lambda \subset \mathbb{L}_1\cup \mathbb{L}_2$, and
$\chi_\Lambda =\sum\limits_{x\in \Lambda}\chi_x$ be the
multiplication by the characteristic function of $\Lambda$. Then,
${\rm d}\Gamma (\chi_\Lambda )=\sum\limits_{x\in
\Lambda}a^*_xa_x=:N_\Lambda$ is the number of particles in
$\Lambda$. The flux of particles out of $\Lambda$ is
\[ I_\Lambda  :=\sum\limits_{(x,y)\in \partial \Lambda}-{\rm
i}h_{x,y}(a^*_xa_y-a^*_ya_x),
\]
where the $h$-boundary of $\Lambda$, $\partial \Lambda$, is the set
of ordered pairs $(x,y)$ of sites $x,y\in \mathbb{L}_1\cup
\mathbb{L}_2$ such that $h_{x,y}\neq 0$, $x\in \Lambda$, $y\in
\mathbb{L}_1\cup \mathbb{L}_2\setminus \Lambda$.

If $\Lambda$ is finite, then $<\alpha _t(N_\Lambda )>_{\rm stat}$ is
finite and independent of $t$, therefore $J_\Lambda :=<I_\Lambda
>_{\rm stat}=\sum\limits_{(x,y)\in
\partial \Lambda}j_{x,y}=0$.

If $\Lambda$ is infinite, then $<N_\Lambda
>_0$ is infinite, whenever the particle-density is
non-zero in the initial state, i.e. $\mu _i>0$ in Eq. (\ref{2.4}).
The same holds for the evolved states $<\cdot >_t$ and their limit
$<\cdot
>_{\rm stat}$.

We consider below the case $\Lambda =\mathbb{L}_1$, whereby we
suppose $\mu _1>0$. Then, $N_{\mathbb{L}_1}$ represents the number
of particles in the first reservoir, which is a conserved quantity
for $\alpha _t^0$, i.e. $[N_{\mathbb{L}_1},{\rm d}\Gamma (h^0)]=0$.
However, in this case, $\partial \mathbb{L}_1=\left\{ \{ x,y\};\;
x\in S_1,\, y\in S_2\right\}$ is finite, so the stationary current
from the first reservoir $J_{\mathbb{L}_1}=<I_{\mathbb{L}_1}>_{\rm
stat}$ makes sense and equals
\begin{equation}\label{2.21}
    J_{\mathbb{L}_1}=\sum\limits_{x\in S_1,\, y\in S_2}j_{x,y},
\end{equation}
whereby $h_{x,y}=v_{x,y}$. It is worth mentioning that, for any
finite sets $\Lambda _1, \Lambda _2$, such that $S_1\subset \Lambda
_1\subset {\mathbb{L}_1}$ and $S_2\subset \Lambda _2\subset
{\mathbb{L}_2}$, the following equalities hold:
$$J_{\mathbb{L}_1\setminus \Lambda _1}= J_{\mathbb{L}_1\cup \Lambda _2}= J_{\mathbb{L}_1},$$
showing that the same current traverses any finite contour
surrounding $S_1$ in $\mathbb{L}_1$, or surrounding $S_2$ in
$\mathbb{L}_2$.

We shall calculate below $j_{x,y}$, Eq.(\ref{2.20}), for $x\in
S_1,\, y\in S_2$, using Eq. (\ref{2.18}). One has
\[     (W_-^*\delta _x,E(k)W_-^*\delta _y)
     =(\delta _x,[I-r^0_+Q_+]E(k)[I-Q_-r^0_-]\delta
     _y) ,
\]
where $Q_\pm ,r^0_\pm$ are calculated at $e=\omega (k)$. Using that
both $r^0_\pm (k)$ and $E(k)$ have block-diagonal structure with
respect to the reservoirs, one obtains
\begin{equation}\label{2.22}
\begin{array}{r}
j_{x,y}=2v_{x,y}\int\limits_{\mathbb{T}_1}{\rm d}k\rho
^0_1\Im{\left[ -E_1(k)Q_-^{(1,2)}r^0_{2,-}+r^0_{1,+}Q_+^{(1,1)}
     E_1(k)Q_-^{(1,2)}r^0_{2,-}\right] _{x,y}} \\ +2v_{x,y}\int\limits_{\mathbb{T}_2}{\rm d}k\rho
^0_2\Im{\left[ -r^0_{1,+}Q_+^{(1,2)}E_2(k)+r^0_{1,+}Q_+^{(1,2)}
     E_2(k)Q_-^{(2,2)}r^0_{2,-}\right] _{x,y}}.\end{array}
\end{equation}
Taking advantage of the fact that all factors, but $E_i(k)$, under
the integral signs depend solely on $e=\omega (k)$, one can perform
the integrals over the energy shells $\Sigma _i(e)$ as in Eq.
(\ref{2.13}):

\begin{equation}\label{2.23}
\begin{array}{l}
    j_{x,y}= \\ 2v_{yx} \int\limits_0^{2d_1}{\rm d}e\rho ^0_1(e)\Im{\left[
    -P_1(e)Q_-^{(1,2)}(e)r^0_{2,-}(e)+r^0_{1,+}(e)Q_+^{(1,1)}(e)
     P_1(e)Q_-^{(1,2)}(e)r^0_{2,-}(e)\right] _{x,y}} \\  +2v_{yx} \int\limits_0^{2d_2}{\rm d}e\rho
    ^0_2(e)\Im{\left[
    -r^0_{1,+}(e)Q_+^{(1,2)}(e)P_2(e)+r^0_{1,+}(e)Q_+^{(1,2)}(e)
     P_2(e)Q_-^{(2,2)}(e)r^0_{2,-}(e)\right]  _{x,y} }
     \end{array}
\end{equation}

The summation over $x,y$ in Eq.(\ref{2.23}) allows a significant
simplification of the expression for the total current:

\[ \begin{array}{rl}
J_{\mathbb{L}_1}=  & 2\Im{\int\limits_0^{2d_1}{\rm d}e \rho
^0_1(e){\rm
    tr}_1\left( -P_1(e)Q_-^{(1,2)}(e)r^0_{2,-}(e)v+r^0_{1,+}(e)Q_+^{(1,1)}(e)
     P_1(e)Q_-^{(1,2)}(e)r^0_{2,-}(e)v\right) } \\ &  +2\Im{ \int\limits_0^{2d_2}{\rm d}e
    \rho ^0_2(e) {\rm
    tr}_2\left( -vr^0_{1,+}(e)Q_+^{(1,2)}(e)P_2(e)+vr^0_{1,+}(e)Q_+^{(1,2)}(e)
     P_2(e)Q_-^{(2,2)}(e)r^0_{2,-}(e)\right)  }  \\ =  & 2\Im{\int\limits_0^{2d_1}{\rm d}e \rho
^0_1(e){\rm
    tr}_1\left( P_1(e)Q_-^{(1,1)}(e)-P_1(e)Q_-^{(1,1)}(e)r^0_{1,+}(e)Q_+^{(1,1)}(e)
     \right) } \\ & +2\Im{\int\limits_0^{2d_2}{\rm d}e
    \rho ^0_2(e){\rm
    tr}_2\left( Q_+^{(2,2)}(e)P_2(e)-
     Q_-^{(2,2)}(e)r^0_{2,-}(e)Q_+^{(2,2)}(e)P_2(e)\right) },
    \end{array}\]
where we applied the identity $Q(z)r^0(z)v=vr^0(z)Q(z)=v-Q(z)$ and
the cyclic invariance of the trace. Now, $2\Im{{\rm tr}}A=-{\rm i\,
tr}(A-A^*)$ and $Q_-^{(i,i)}-Q_+^{(i,i)}=\sum
_jQ_+^{(i,j)}(r_{j,+}-r_{j,-})Q_-^{(j,i)}=\sum _jQ_+^{(i,j)}2{\rm
i}\pi P_jQ_-^{(j,i)}$, so
\begin{equation}\label{2.24}
    J_{\mathbb{L}_1}=2\pi \int\limits_0^{2\min{(d_1,d_2)}}{\rm d}e(\rho ^0_1(e)-\rho ^0_2(e)){\rm
    tr}_1\left( P_1(e)Q_-^{(1,2)}(e)P_2(e)Q_+^{(2,1)}(e)\right) .
\end{equation}

\subsection{The charge-density profile in the stationary state}\label{s2.3}

The charge density at the site $x\in \mathbb{L}_1\cup \mathbb{L}_2$
in the stationary state is given, according to Eq.(\ref{2.11}), by:
\begin{equation}\label{2.25}
d(x):=<a^*_xa_x>_{\rm stat}=\left( \delta _x,\rho _+\delta _x\right)
= d_{\rm p}(x)+d_{{\rm ac}}(x),
\end{equation}
where we separated the contributions of the point and absolutely
continuous spectra:
\begin{equation}\label{2.26}
\begin{array}{l}
d_{\rm p}(x)= \sum\limits_{\lambda \in \sigma _{\rm p}(h)}\left(
P_\lambda \delta _x,\, \rho ^0P_\lambda \delta _x\right)=
\sum\limits_{i=1,2}\int\limits_{k\in \mathbb{T}_i} {\rm d}kf_{\beta
_i,\mu _i}(\omega (k)) \sum\limits_{\lambda \in \sigma _{\rm
p}(h)}|\left( P_\lambda \phi (k), \delta _x\right) |^2, \\  \\
d_{{\rm ac}}(x)=\left( W_-^*\delta _x,\, \rho ^0W_-^*\delta
_x\right) = \sum\limits_{i=1,2}\int\limits_{k\in \mathbb{T}_i} {\rm
d}kf_{\beta _i,\mu _i}(\omega (k))|\left[ W_-\phi (k)\right] _x|^2.
\end{array}
\end{equation}

Clearly, $\sum\limits_{x\in \mathbb{L}_2}d_{\rm p}(x)<\infty$,
meaning that the number of particles accommodated on the eigenstates
of $h$ is finite and located near $S_1\cup S_2$.

We calculate $d_{{\rm ac}}(x)$ for $x\in \mathbb{L}_2$. Remembering
Eq.(\ref{2.17})  and the definition (\ref{2.5}), we have
\begin{equation}\label{2.27}
|\left[ W_-\phi (k)\right] _x|^2=\left( \delta _x,\, [I-r^0_+(\omega
(k))Q_+(\omega (k))]E(k) [I-Q_-(\omega (k))r^0_-(\omega (k))]\delta
_x\right) .
\end{equation}
Again, the integration over the energy shell $\omega (k)=e$ can be
performed with Eq. (\ref{2.13}), so, finally,
\begin{equation}\label{2.28}
d_{{\rm ac}}(x)=\sum\limits_{i=1,2}\int\limits_0^{2d_i}{\rm
d}ef_{\beta _i,\mu _i}(e)\left( \delta _x,\,
[I-r^0_+(e)Q_+(e)]P_i(e) [I-Q_-(e)r^0_-(e)]\delta _x\right) .
\end{equation}
Here, one should take into account  the block structure of $r^0$ and
the fact that $P_i(e)_{y,z}=0$ unless both $y,z\in  \mathbb{L}_i$.
For instance, if $x\in \mathbb{L}_2$, the term $i=1$ in
Eq.(\ref{2.28}) simplifies to
\begin{equation}\label{2.29}
d^{(1)}(x)=\int\limits_0^{2d_1}{\rm d}ef_{\beta _1,\mu _1}(e)
\sum\limits_{s_2,s'_2\in
S_2}g_{2,+}(e;s_2-x)[Q_+^{(2,1)}(e)P_1(e)Q_-^{(1,2)}(e)]_{s_2,s'_2}g_{2,-}(e;x-s'_2),
\end{equation}
while the term $i=2$ has a more complicated structure, due to the
superposition of the incident and the reflected waves:
\begin{equation}\label{2.30}
\begin{array}{rl}
d^{(2)}(x)= & \int\limits_0^{2d_2}{\rm d}ef_{\beta _2,\mu _2}(e)\{  P_2(e)_{x,x} \\
-  & \sum\limits_{s_2\in S_2}\left[g_{2,+}(e;x-s_2)\left(
Q_+^{(2,2)}(e)P_2(e)\right) _{s_2,x}+\left(
P_2(e)Q_-^{(2,2)}(e)\right) _{x,s_2}g_{2,-}(e;s_2-x) \right]
\\ + &\sum\limits_{s_2,s'_2\in S_2} g_{2,+}(e;x-s_2)[Q_+^{(2,2)}(e)P_2(e)Q_-^{(2,2)}(e)]_{s_2,s'_2}g_{2,-}(e;s'_2-x)\} .
\end{array}
\end{equation}

The following remark is in order. If the dimension $d_2\geq 2$, the
Green function $g_{2,\pm} (e;x)\to 0$ as $x\to \infty$. Hence, as
$x\to \infty$ in $\mathbb{L}_2$, $d(x)$ approaches the constant
$\rho _2= \int\limits_0^{2d_2}{\rm d}ef_{\beta _2,\mu _2}(e)
P_2(e)_{x,x}$, which is the initial equilibrium density of the
second reservoir. Indeed, turning on a tunneling junction should not
affect the intensive parameters defining the reservoirs, this is one
of the reasons why we restrict to dimensions of the lattices larger
than 1.

\section{Numerical computations}\label{s3}
\setcounter{equation}{0}

In the computations we particularize to two-dimensional homogeneous
reservoirs, i.e. we take $\mathbb{L}_1,\, \mathbb{L}_2$ both equal
to $\mathbb{Z}^2$, and $h^0_1,\; h^0_2$ both equal to $-(1/2)\Delta$
acting in $l^2(\mathbb{Z}^2)$ , see Eq.(\ref{2.1}). Hence,
$\mathbb{T}_1=\mathbb{T}_2=\mathbb{T}=(-\pi ,\pi )^2$ and the
dispersion law of $h^0$ is $\omega (k_1,k_2)=\sum\limits_{\alpha
=1}^2 2\sin ^2{(k_\alpha /2)}$.

\subsection{Evaluation of the Green's function of the 2D Laplacian}\label{s3.1}

We consider here the function $g(z;x)=\left[ (-(1/2)\Delta
-z)^{-1}\right] _{0,x}$. It is defined, for $z$ in the complex plane
cut along the real segment $[0,\; 4]$ and for $x=(x_1,\; x_2) \in
\mathbb{Z}^2$, by the formula (\ref{2.12}):
\begin{equation}\label{3.1}
    g(z;x_1,x_2)=(2\pi )^{-2}\int\limits_{\mathbb{T}}{\rm d}k_1{\rm
    d}k_2 \frac{{\rm e}^{-{\rm i}(k_1x_1+k_2x_2)}}{\omega (k_1,k_2)-z}.
\end{equation}

\begin{Remark}
As done in Eq.(\ref{2.12}), we can integrate first over the energy
shells. For $e<2$, the energy shell  $\Sigma (e)=\{ k=(k_1,k_2)\in
\mathbb{T};\; \omega (k)=e\} $ is a closed curve, which can be
parametrized by one angle $\theta \in [0,2\pi )$. Indeed, for all
$e\in (0,2)$ and any $\theta$, there exists one point $k(e,\theta
)\in \Sigma (e)$ along that direction. Denoting $K(e,\theta )$ the
norm of this $k$, we have
\[ \left\{ \begin{array}{c} k_1(e,\theta )=K(e,\theta )\cos{\theta } \\ k_2(e,\theta )=K(e,\theta )\sin{\theta }
\end{array}\right. ,
\]
with $K(e,\theta )$ the unique solution of the equation $\omega
(K\cos{\theta},K\sin{\theta})=e$. The change of variables
$(k_1,k_2)\to (e, \theta )$ is regular from $|k_1|+|k_2|<\pi$ to
$(0,2)\times [0,2\pi)$ and its Jacobian equals
\begin{equation}\label{3.2}
    J(e,\theta )=K(e,\theta )\partial _eK(e,\theta ).
\end{equation}
Hence, if $e<2$, and denoting $\phi$ the polar angle of
$x=(x_1,x_2)$, we have
\begin{equation}\label{3.3}
P(e)_{0,x}=\frac{1}{(2\pi )^2}\int\limits_0^{2\pi}{\rm d}\theta
J(e,\theta )\exp{[-{\rm i}|x|K(e,\theta )\cos{(\phi -\theta )}]}.
\end{equation}
For $e>2$, $\Sigma (e)$ consists of four disconnected parts, which
can be put together by changes of variables of the form $k'_\alpha
=k_\alpha \pm \pi$ into the closed curve corresponding to $\Sigma
(4-e)$ {\rm (see Figure \ref{f1})}, which can be parametrized by an angle
$\theta$ as before.
\begin{figure}
\hskip 4cm\includegraphics{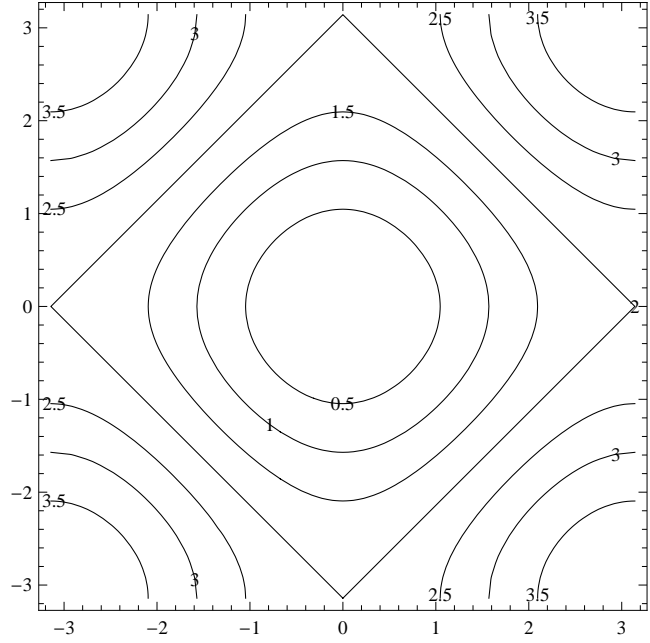} \caption{Energy levels
$\Sigma (e)$}\label{f1}
\end{figure}
\end{Remark}

\begin{figure}
\hskip 3cm\includegraphics{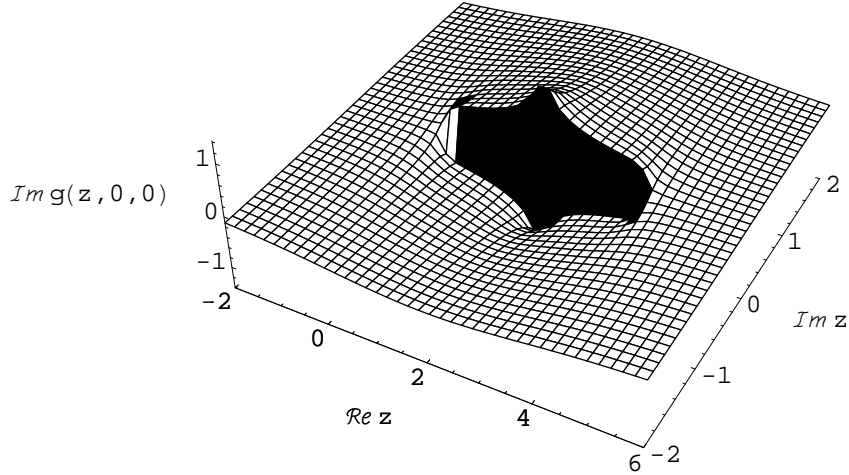} \caption{Graph of
$\Im{g(z;0,0)}$ in the complex $z$-plane cut along $[0,4]$}\label{f2}
\end{figure}

One can easily see that
\begin{equation}\label{3.4}
    g(z;x_1,x_2)=g(z;x_2,x_1)=g(z;|x_1|,|x_2|)=(-1)^{1+x_1+x_2}g(4-z;x_1,x_2)=\overline{g({\bar
    z};x_1,x_2)}.
\end{equation}
Hence, the values of $g(z;x)$ are real for real $z$ outside the
segment $[0,\; 4]$. We plotted in Figure \ref{f2}, as an example,
$\Im{g(z;0,0)}$, which is strictly positive on the upper part of the
cut. Also, $g$ is determined by the values $g(z;m,n)$ with $0\leq
m\leq n$ and $\Re{z}\leq 2)$.

Using the identity $A^{-1}={\rm i}\int _0^\infty {\rm e}^{-{\rm
i}At}{\rm d}t,\;$ if $\;\Im{A}<0$, $g$ can be expressed, for $z$ in
the upper complex half-plane, as an integral of a product of two
Bessel functions of integer index,
\begin{equation}\label{3.5}
    g(z;m,n)=\mathrm{i}^{m+n+1}\int_{0}^{\infty}{\rm
d}te^{-\mathrm{i}t(2-z)}J_m(t)J_n(t),
\end{equation}
where we used the representation
\[
J_n(t)=\frac{(-\mathrm{i})^{n}}{2\pi}\int_{-\pi}^{\pi}\exp(-\mathrm{i}nk
+\mathrm{i}t\cos{k}){\rm d}k .
\]
The integral (\ref{3.5}) is related to a regularized hypergeometric
function \cite{BE}:
\begin{equation}\label{3.6}
g(z;m,n)=
\left(-\frac{1}{2}\right)^{m+n+1}\frac{1}{(z-2){}^{m+n+1}}(m+n)!^{2}\cdot
{}_{4}F_{3}^{reg}(a_1,a_2,a_3,a_4;b_1,b_2,b_3;\frac{4}{(z-2)^{2}}),
\end{equation}
where $$a_1=a_2=\frac{1+m+n}{2},\; a_3=a_4=\frac{2+m+n}{2},\;
b_1=1+m,\; b_2=1+n,\; b_3=1+m+n.$$ ${}_{4}F_{3}^{reg}(...;...;u)$ is
analytic in the complex-$u$-plane cut along $[1,\, \infty )$. Note
that the limit values $g(e+\textrm{i}0;...)$ are expressed in terms
of $\, {}_{4}F_{3}^{reg}(...;...;\frac{4}{(e-2)^{2}}-\textrm{i}0)$
if $0<e<2$.

For $m,n$ not too large (say, $m+n\leq 5$), Eq. (\ref{3.6})
(including the limit values $g(e+{\rm i}0;m,n)$ for $0<e<4$) can be
safely computed using the built-in functions of Mathematica.
For larger values of $m+n$, an asymptotic study  of $g$
has been performed, based on the steepest descents method
\cite{Erd}, \cite{v}, the output of which is described below:

\bigskip
For $z$ outside $[0,\, 4]$, $g(z;x)$ is the Fourier transform of a
$C^\infty$, periodic function of $k_1,k_2$, hence it decays faster
than any inverse power of the distance $|x|$.

When $z=e+{\rm i}\epsilon$,  $0<e<4$, the denominator in Eq.
(\ref{3.1}) develops a singularity in the integration domain when
$\epsilon \searrow 0$. As remarked below Eq. (\ref{3.4}), it is
sufficient to consider $e<2$ and $x=(m,n),\; m,n\geq 0$. Let $\chi
(t)$ be a $C^\infty$-function, which equals $1$ if $|t-e|<a$ and
vanishes if $|t-e|>b$, where $0<a<b<2-e$. Then, the asymptotic
series of $g(e+{\rm i}0;x)$ coincides with that of
\begin{equation}\label{3.7}
g_\chi (e+{\rm i}0;x)=(2\pi )^{-2}\int\limits_{\mathbb{T}}{\rm
d}k_1{\rm
    d}k_2 \frac{\chi (2-e-\cos{k_1}-\cos{k_2}){\rm e}^{-{\rm i}(k_1m+k_2n)}}{2-e-{\rm i}0-\cos{k_1}-\cos{k_2}}.
\end{equation}

Therefore, remembering Eq.(\ref {3.3}),
\begin{equation}\label{3.8}
    g_\chi (e+{\rm i}0;x)=\int {\rm d}t\frac{\chi (t-e)}{t-e-{\rm i}0} P(t)_{0,x}.
\end{equation}

The phase $\Phi (e,\phi ;\theta )=-K(e,\theta )\cos{(\phi -\theta
)}$ in Eq. (\ref{3.3}) (where $\phi =\arctan{(n/m)}$ is the
direction of $x=(m,n)$) has two stationary points with respect to
$\theta$, i.e. the equation $\partial _\theta \Phi (e,\phi;\theta
)=0$ has two solutions $\theta _s(e,\phi)\in [0,\pi /2]$ and $
\theta _s(e,\phi)+\pi$, which control the asymptotic behavior.
Thereby,
\begin{equation}\label{3.9}
    \partial _\theta ^2\Phi (e,\phi ;\theta )|_{\theta _s}=J(e,\theta _s)^2\psi (e,\phi   )>0,
\end{equation}
where
\begin{equation}\label{3.10}
    \psi (e,\phi )=\cos{(K_s\cos{\theta _s})}\sin{(K_s\sin{\theta _s})}\sin{\phi}+
    \cos{(K_s\sin{\theta _s})}\sin{(K_s\cos{\theta
    _s})}\cos{\phi},
\end{equation}
Here we denoted, for shortness, $K_s=K(e,\theta _s(e,\phi )),\;
\theta _s=\theta _s(e,\phi)$. Likewise, $\partial _\theta ^2\Phi
(e,\phi ;\theta )|_{\theta _s+\pi}<0$. Also, $\partial _e\Phi
(e,\phi ; \theta _s)=-\partial _eK(e,\theta _s)\cos{(\phi -\theta
_s)}<0$, as $\partial _eK>0$ and $\phi -\theta _s\in [-\pi /2,\, \pi
/2]$.

A straightforward application of Lemma 7.2.5 of \cite{v} shows that
the leading term of the $|x|\to \infty$ asymptotical series of Eq.
(\ref{3.8}) equals
\begin{equation}\label{3.11}
    g(e+{\rm i}0;x)\sim \frac{1}{\sqrt{|x|}
    \sqrt{2\pi \psi (e,\phi )}}
    e^{\textrm{i}(|x|K_s\cos{(\phi -\theta _s)}+\pi /4)}.
\end{equation}

\subsection{The density profile}\label{s3.2}

We compute here the density distribution (i.e. the number of band
electrons per lattice site) in the second reservoir $\mathbb{L}_2$
in the stationary state. Thereby, we take the initial equilibria to
correspond to zero temperature in both reservoirs, and to chemical
potentials $\mu _1$ in $\mathbb{L}_1 $, and $\mu _2<\mu _1$ in
$\mathbb{L}_2$. Hence, the initial (equilibrium) densities in the
reservoirs take the values:
\begin{equation}\label{3.12}
\rho ^{(i)}_{\rm eq}=\int\limits_0^{\mu _i}P(e)_{0,0}{\rm d}e.
\end{equation}
We choose for exemplification $\mu _1=1.4$, $\mu _2=0.3$, what
corresponds to $\rho ^{(1)}_{\rm eq}=0.2804$, $\rho ^{(2)}_{\rm
eq}=0.0492$.

Moreover, $S_i=\{ s^{i,1},\, s^{i,2}\} ,\; (i=1,2)$ consist of two
points at distance $d_i$ apart along an axis of $\mathbb{L}_i$, e.g.
$s^{i,1}=(0,0)$ and $s^{i,2}=(d_i,0) $. The "contacts" between
reservoirs is done by direct tunneling between the two pairs of
points, $\{ s^{1,j},\, s^{2,j}\}$ with tunneling constants $t_j$,
($j=1,2$). Thereby, we fix $d_2=20$ and $t_2=1$, what proves to be
good for a nice visualization of the interference phenomena, while
$d_1$ and $t_1$ are left as parameters.

Under the assumptions made above, the transmitted and reflected
densities in the second reservoir, Eqs. (\ref{2.29}) and
(\ref{2.30}), acquire the form (for simplicity, we do not exhibit
the dependence on $d_1$ and $t_1$):
\begin{equation}\label{3.13}
d^{(i)}(x)=\int\limits_0^{\mu_i}\delta ^{(i)}(e;x){\rm
d}e\;\;(i=1,2),\; x\in \mathbb{L}_2,
\end{equation}
where the functions under the integral sign have the following
structure:
\begin{description}
  \item[$i=1$ (transmitted)]
\begin{equation}\label{3.14}
\delta ^{(1)}(e;x)=\left( V_-(e,x), m_{\rm tr}(e)V_-(e,x)\right)
_{\mathbb{C}^2},
\end{equation}
where
\begin{equation}\label{3.15}
m_{\rm tr}(e)=Q_+^{(2,1)}(e)\cdot (P(e)|_{S_1\times S_1})\cdot
Q_-^{(1,2)}(e)>0,
\end{equation}
with $P(e)|_{S_1\times S_1}$ denoting the $2\times 2$ matrix $\{
P(e)_{s_1,s'_1},\; s_1,s'_1\in S_1\}$ and $V_-(e,x)$ denoting the
2-dimensional vector $\{ g_-(e;x-s_2),\; s_2\in S_2\}$; the matrices
$Q_\pm ^{(2,1)}(e)_{s_2,s_1}$ were defined after Eq.(\ref{2.16}).
Clearly, the whole dependence on $x$ is contained in $V_-(e;x)$,
whose components are "waves" of the same shape (independent of
$t_1,d_1$) originating at the two points of $S_2$. The shape of
these waves is given, far from the source, by the levels of the
function at the exponent of Eq. (\ref{3.11}).

The matrix $m_{\rm tr}(e)$, depending on the details of the
interaction $v$ (in particular on $d_1,t_1$), controls the
interference of the two waves. Indeed, Eq. (\ref{3.14}) describes
the squared norm of the linear combination $m_{\rm
tr}(e)^{1/2}V_-(e;x)$.

  \item[$i=2$ (reflected)]
  \begin{equation}\label{3.16}
\delta ^{(2)}(e;x)=P(e)_{0,0}-2\Re{\left(
V_-(e,x),Q_+^{(2,2)}(e)V^{(0)}(e;x)\right) _{\mathbb{C}^2}} + \left(
V_-(e,x),m_{\rm ref}(e) V_-(e,x)\right) _{\mathbb{C}^2},
\end{equation}
where
\begin{equation}\label{3.17}
m_{\rm ref}(e)=Q_+^{(2,2)}(e)\cdot (P(e)|_{S_2\times S_2})\cdot
Q_-^{(2,2)}(e)>0,
\end{equation}
with $V^{(0)}(e;x)$ denoting the 2-dimensional vector $\{
P(e)_{x,s_2},\; s_2\in S_2\}$. The first term in Eq. (\ref{3.16}) is
a constant, giving the density per energy and site of "free"
electrons of energy $e$ in the initial equilibrium state of the
reservoir $\mathbb{L}_2$. The last term has the same significance as
for Eq. (\ref{3.14}) with $m_{\rm ref}$ instead of $m_{\rm tr}$. The
middle term is new and gives account of the overlapping between the
"waves" originating in $S_2$ with the free-electron states in
$\mathbb{L}_2$ of energy $e$.
\end{description}

\begin{figure}[h]
\begin{center}
a\includegraphics[width=5cm]{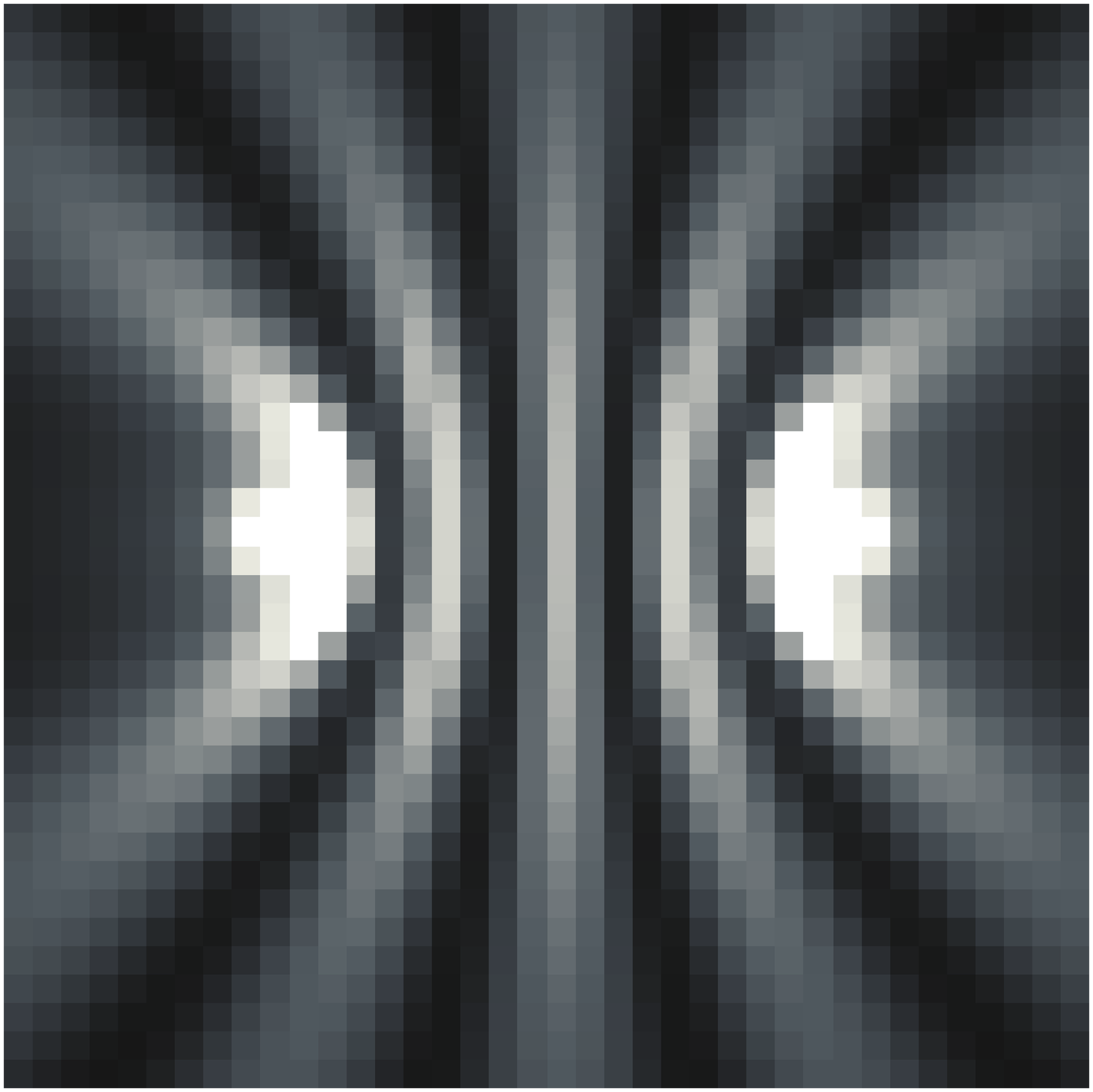}\
b\includegraphics[width=5cm]{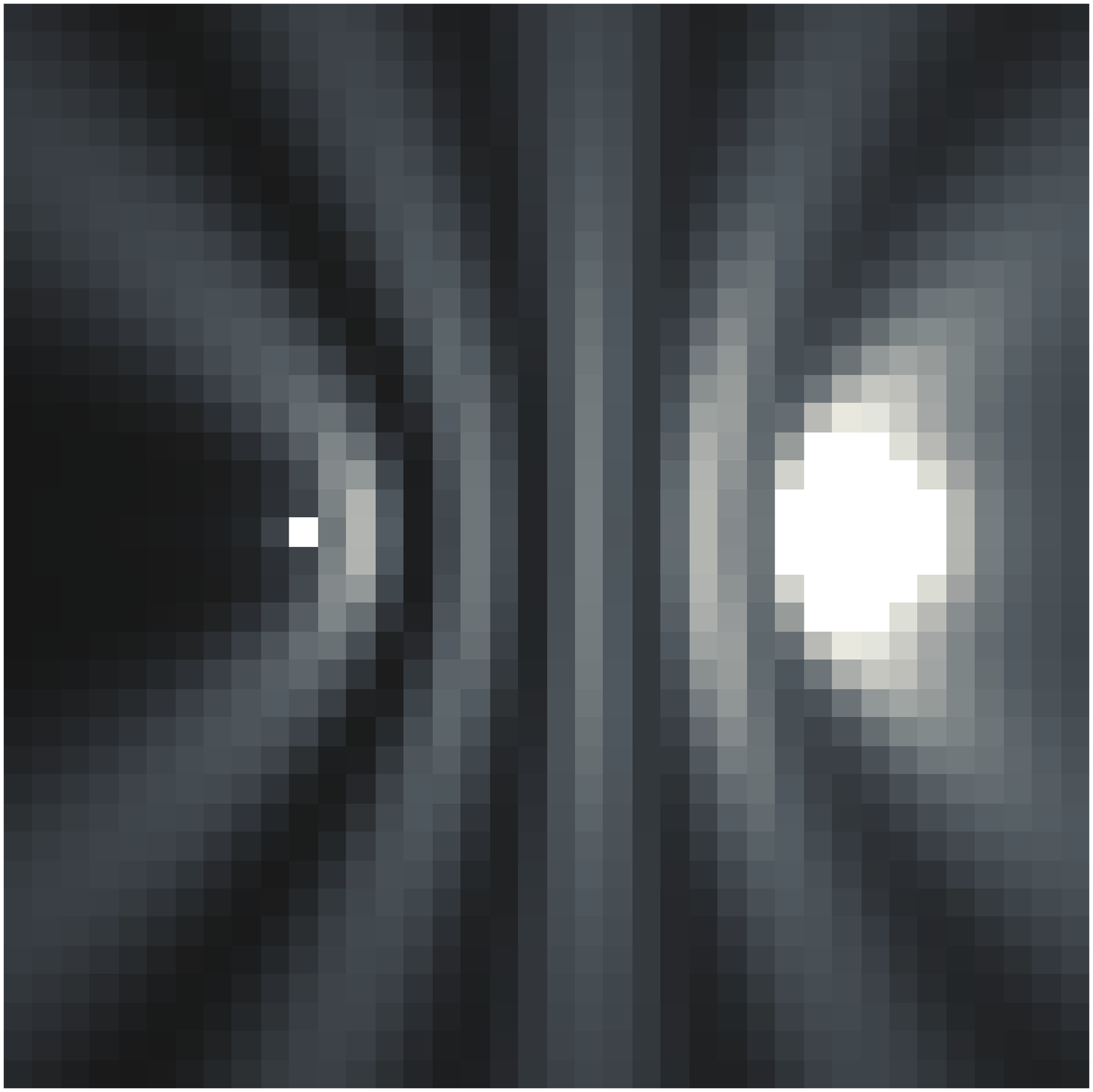}\
c\includegraphics[width=5cm]{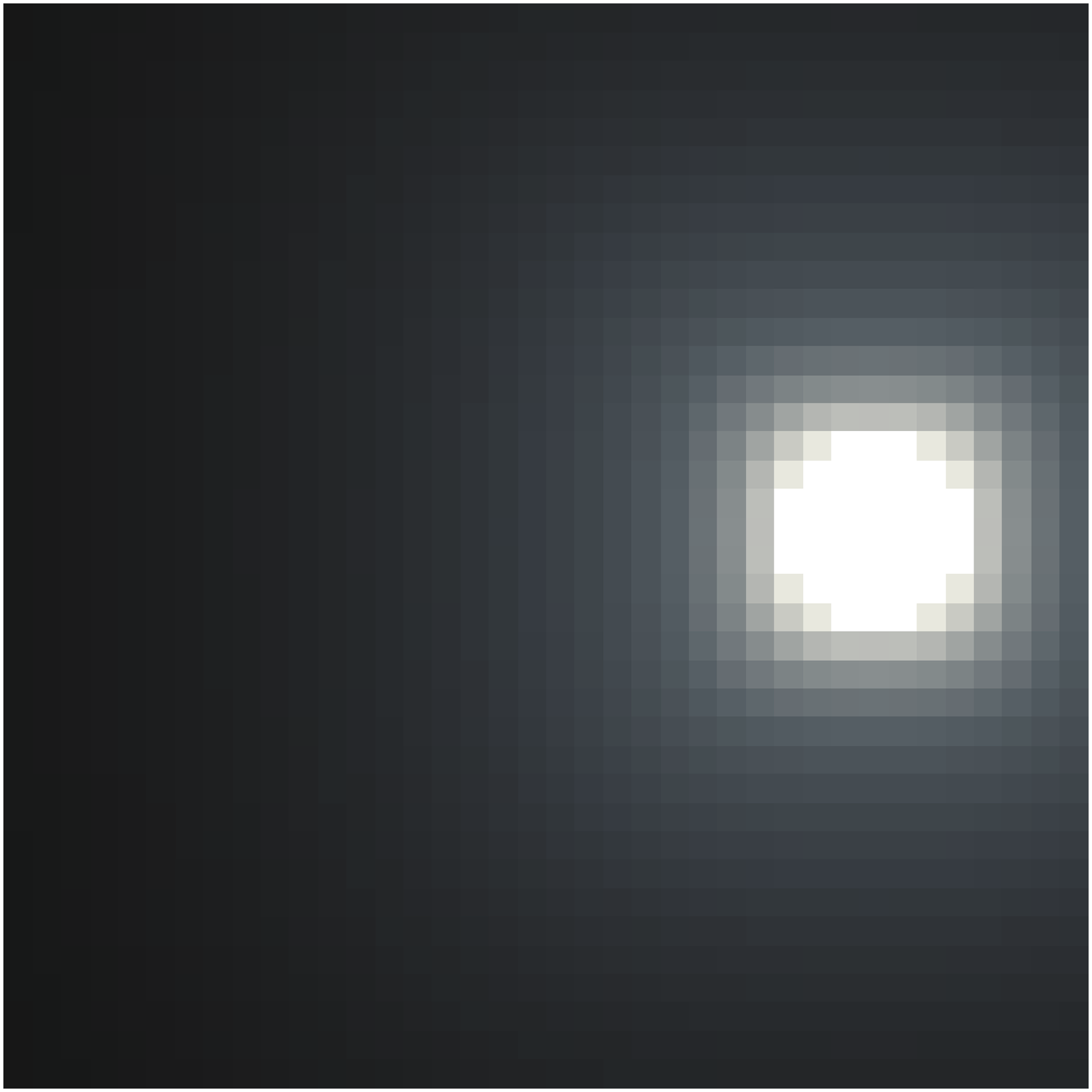}\ \includegraphics[width=0.4cm]{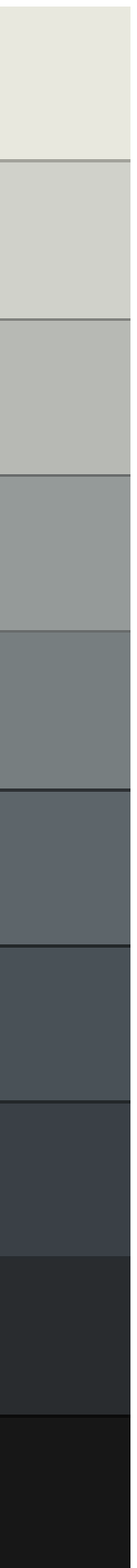}\\
\end{center}
\caption{Density plot of $\delta ^{(1)}(0.3;x)$ for $t_2=1,\, d_1=1$
and: a. $t_1=1$; b. $t_1=1/2$; c. $t_1=0$; Legend values: $k\cdot
10^{-3},\; k=1,...,10$.}\label{f3}
\end{figure}

\begin{figure}[h]
\begin{center}
a\includegraphics[width=5cm]{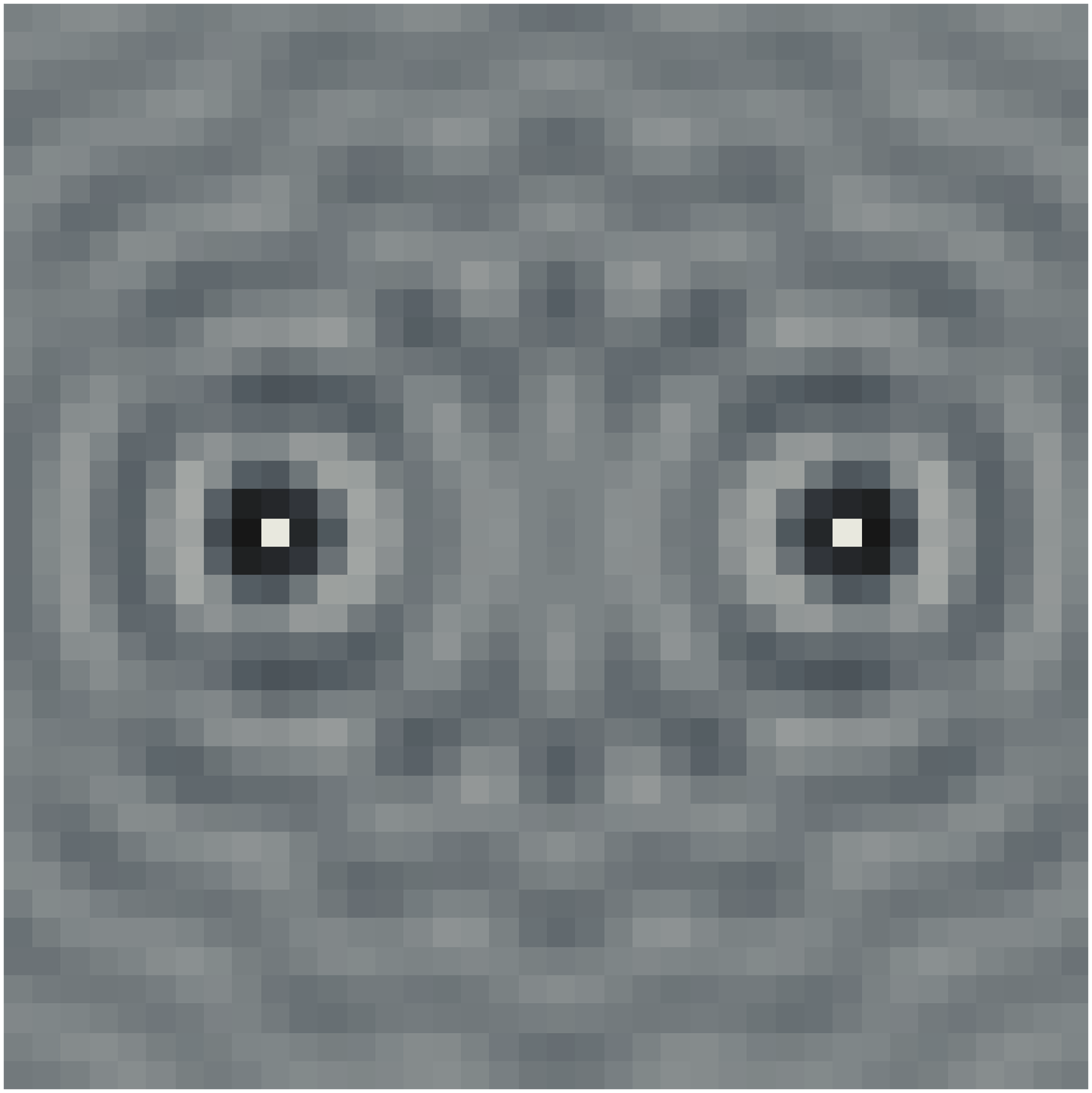}\
b\includegraphics[width=5cm]{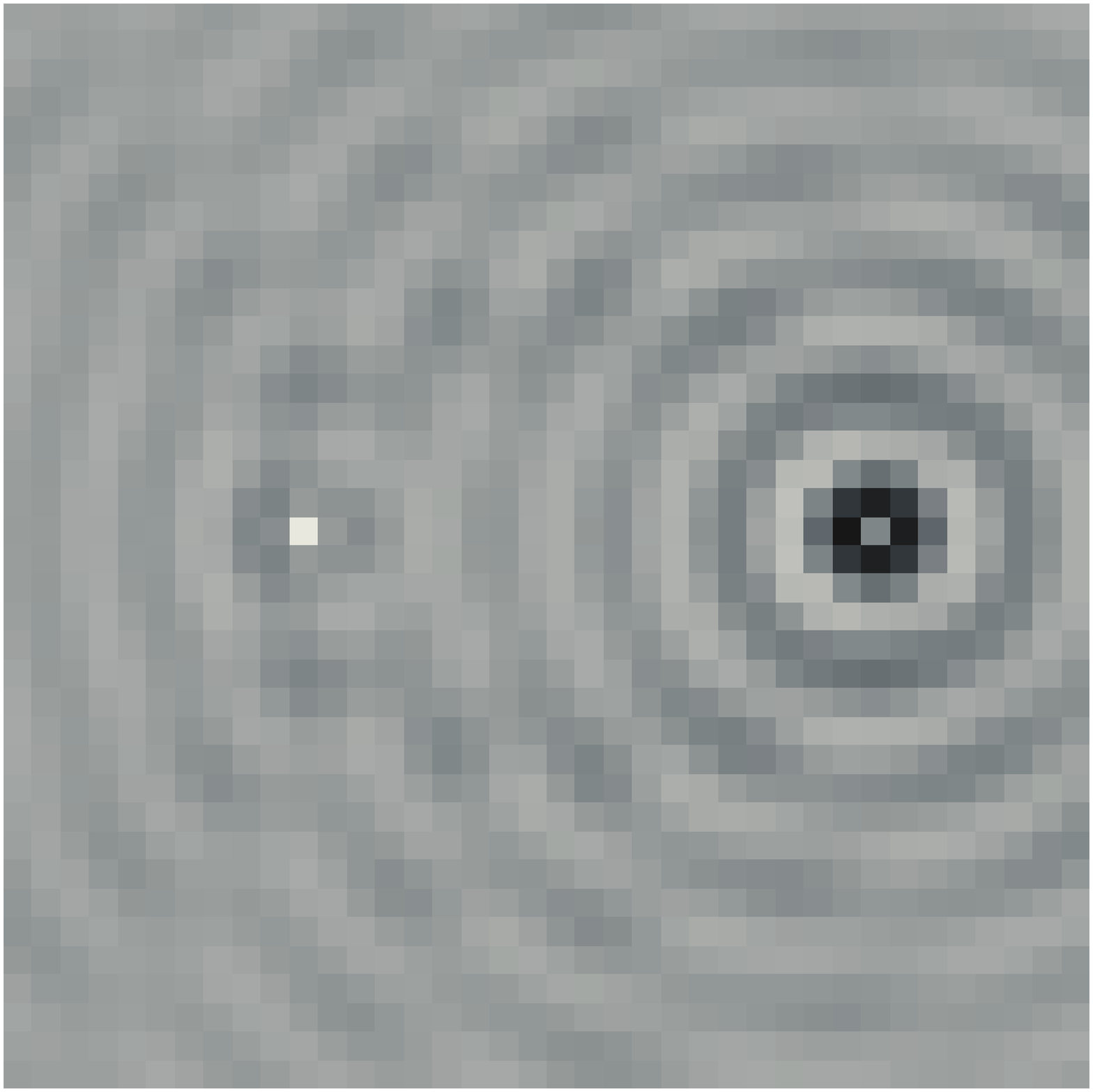}\
c\includegraphics[width=5cm]{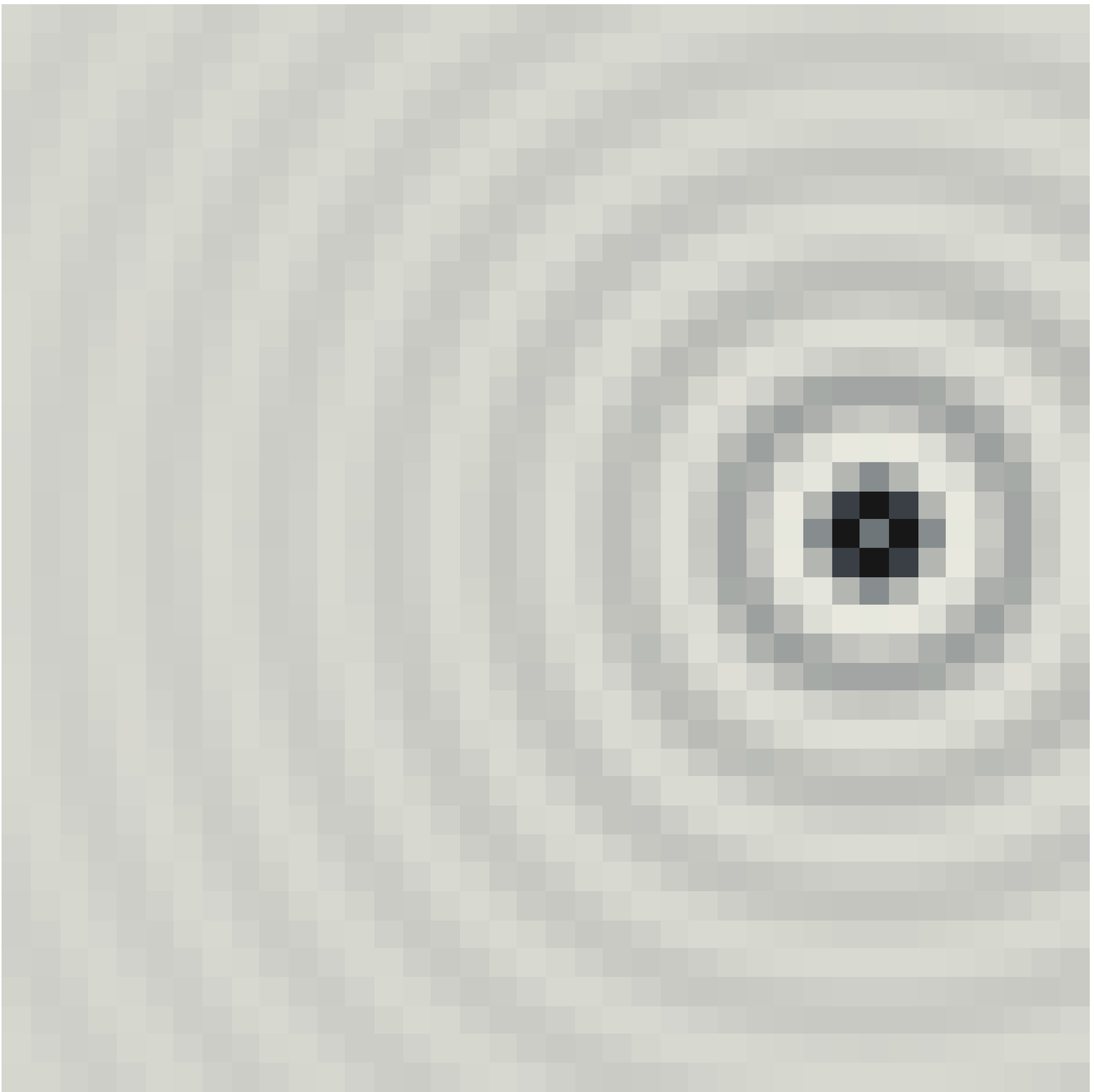}\
\includegraphics[width=0.4cm]{fig4L.eps}\caption{Density plot of
$\delta ^{(2)}(0.3;x)$ for $t_2=1,\, d_1=1$ and: a. $t_1=1$; b.
$t_1=1/2$; c. $t_1=0$. Legend values: $0.095+k\cdot 0.012,\,
k=1,...,10$.}\label{f4}
\end{center}
\end{figure}

\begin{figure}[h]
\begin{center}
a\includegraphics[width=5cm]{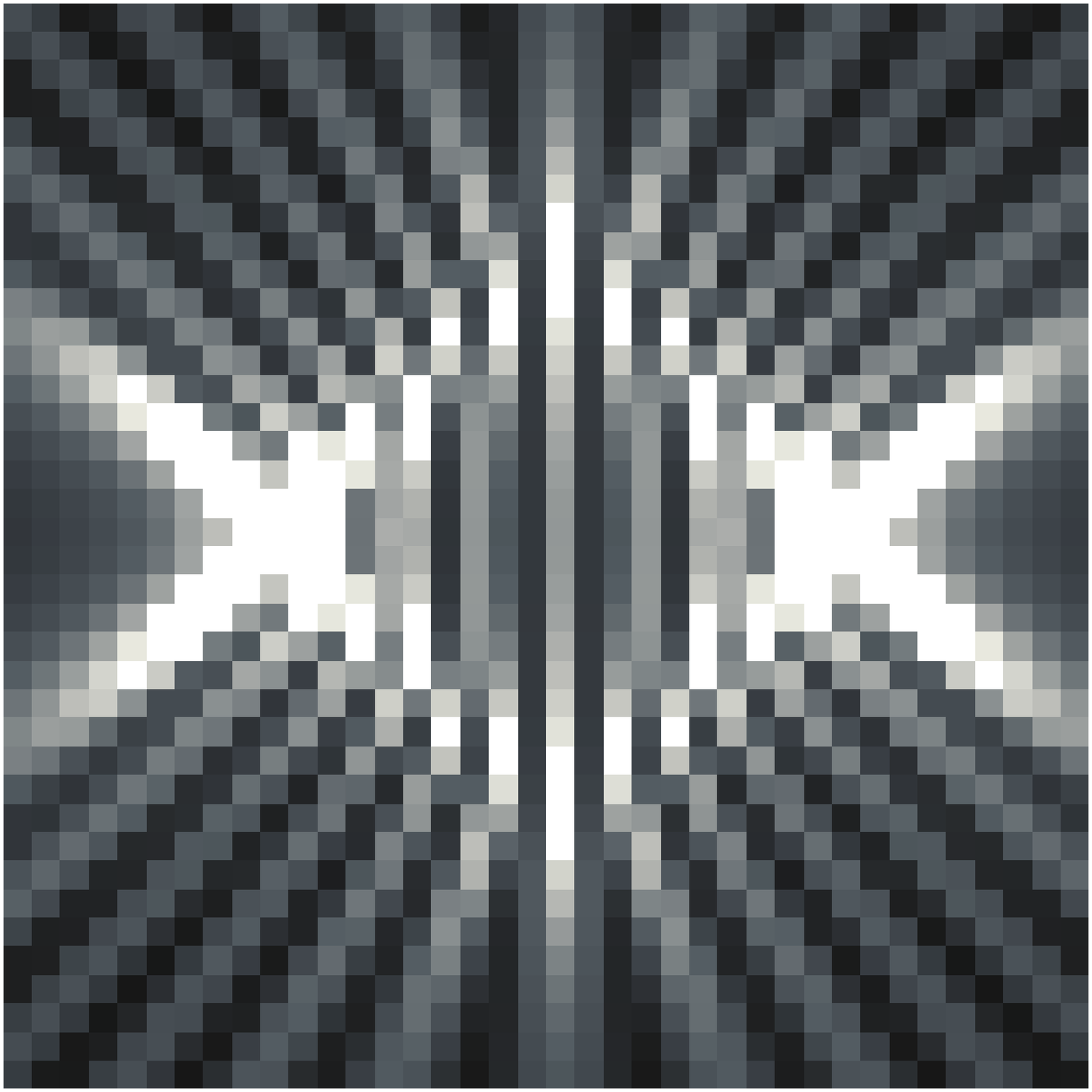}\
b\includegraphics[width=5cm]{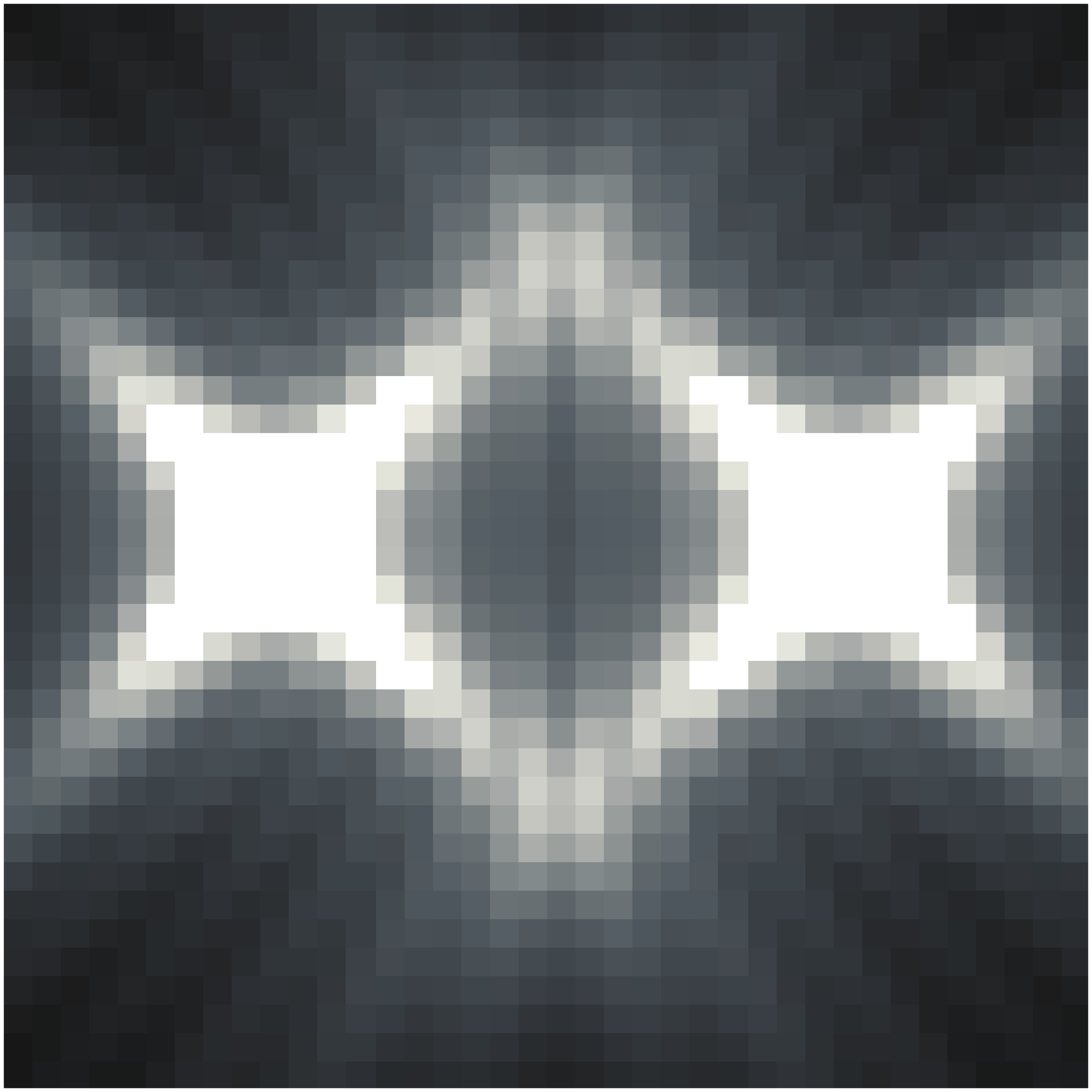}\
c\includegraphics[width=5cm]{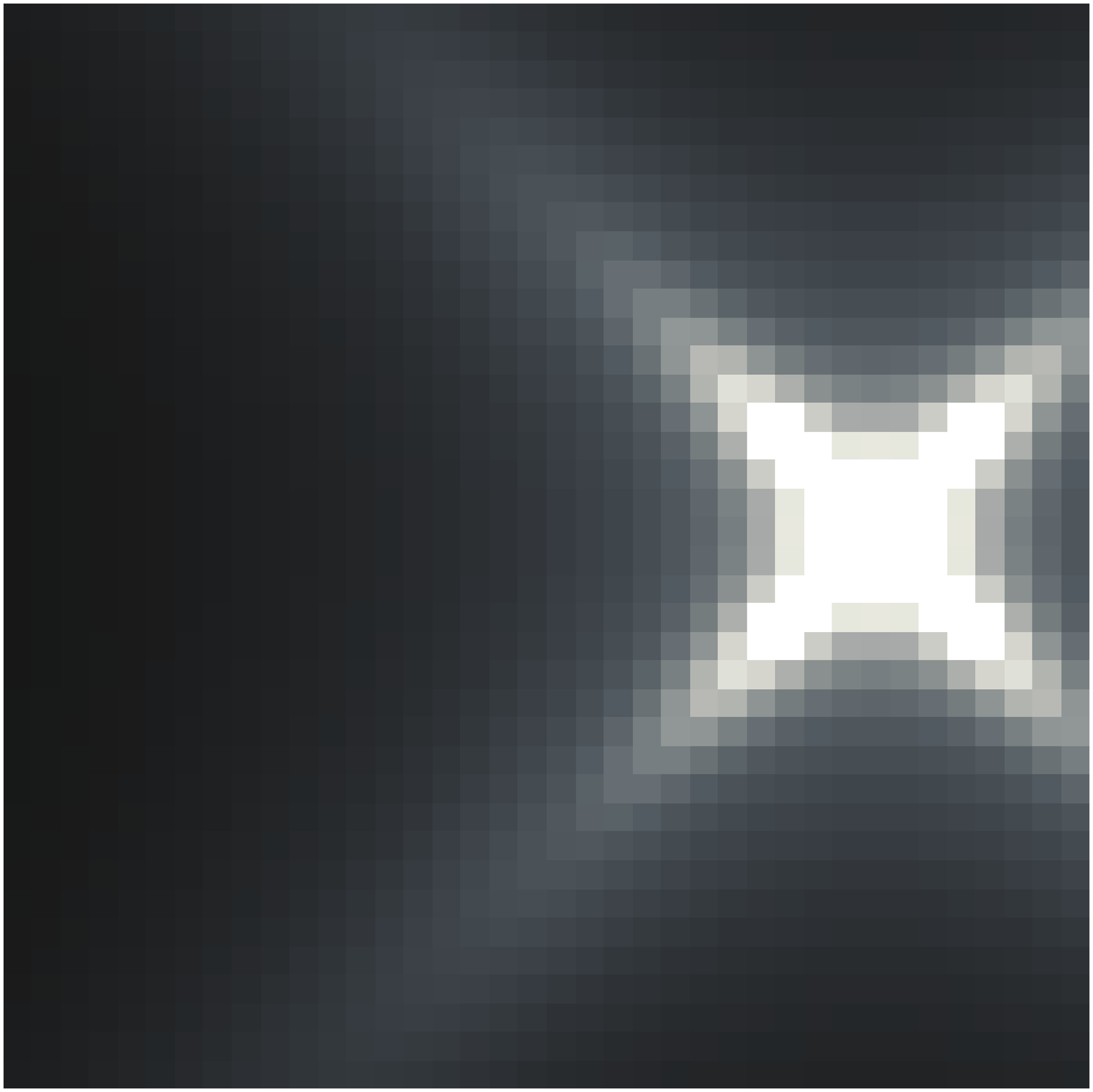}\
\includegraphics[width=0.4cm]{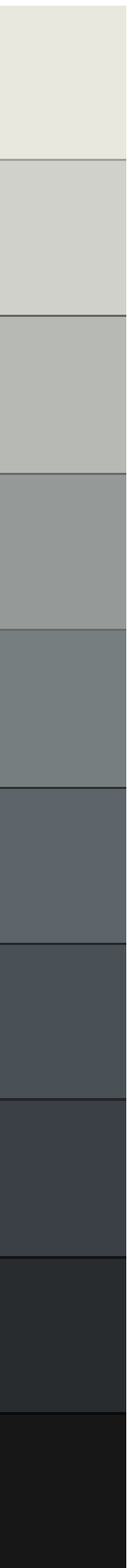}\\
\end{center}
\caption{Density plot of $\delta ^{(1)}(1.4;x)$ for $t_2=1$ and: a.
$t_1=1$, $d_1=1$; b. $t_1=1$, $d_1=20$; c. $t_1=0$.  Legend values:
$6k\cdot 10^{-4},\; k=1,...,10$.}\label{f5}
\end{figure}

We start by exploring the "fixed-energy" density distributions,
$\delta ^{(1)}$ and $\delta ^{(2)}$. In Figures \ref{f3} and
\ref{f4}, density plots of $\delta ^{(1)}(0.3;x)$ and $\delta
^{(2)}(0.3;x)$, respectively, are shown for $x$ in a square of side
40 containing $S_2$. Thereby, the cases $t_1=1$ (symmetric),
$t_1=1/2$ (asymmetric)and $t_1=0$ (corresponding to one contact) are
represented, in order to exemplify the dependence on $t_1$. Figure
\ref{f5} shows the density plot of $\delta ^{(1)}(1.4;x)$ in the
symmetric case $t_1=1$ for a. $d_1=1$ and b. $d_1=20$; panel c is
the plot of $\delta ^{(1)}(1.4;x)$ in the case of one contact
($t_1=0$).

The interference patterns are clearly visible in the symmetric case
and $d_1=1$ (left panels), but the ones for $\delta ^{(1)}$ and for
$\delta ^{(2)}$ are drastically different. The number of fringes
increases with increasing $e$. In the case of $\delta ^{(1)}$, the
contrast of the fringes decreases as $t_1$ decreases, leading, when
$t_1=0$, to their complete disappearance (panels c in Figures
\ref{f5} and \ref{f4}). What concerns $\delta ^{(2)}$, "circular"
fringes around $s_{2,2}$ survive even at $t_1=0$ (Figure \ref{f5}c).

\bigskip
This can be accounted for using the asymptotic form of $V_-(e,x)$,
see Eq. (\ref{3.11}). Indeed, for $\delta ^{(1)}$, the interference
of the two components $g_-(e;x-s_2),\; s_2\in S_2$ in the linear
combination $m_{\rm tr}(e)^{1/2}V_-(e;x)$ is controlled by the phase
difference $|x-s_{2,1}|K_{s,1}\cos{(\phi _1 -\theta
_{s,1})}-|x-s_{2,2}|K_{s,2}\cos{(\phi _2 -\theta _{s,2})}$, where we
denoted with subscripts $1,2$ the corresponding functions calculated
for $s_{2,1},s_{2,2}$, respectively; taking into account that
$K_s\cos{(\phi  -\theta _s)}$ is slowly varying with $\phi$ (e.g. at
$e=0.3$, it oscillates between 0.784 and 0.795, and its derivative
does not exceed 0.022), the above difference is constant (modulo
$2k\pi$) on a family of curves similar to hyperbolae with foci
$s_{2,1},s_{2,2}$. The pattern for $\delta ^{(2)}$ comes from the
second term in Eq. (\ref{3.16}): as $V^{(0)}(e;x)$ is a real vector,
this term is a sum of cosines, $\cos{\left( \psi
_i+|x-s_{2,i}|K_{s,i}\cos{(\phi _i -\theta _{s,i})}\right) }$, which
is constant on curves similar to circles centered at
$s_{2,1},s_{2,2}$. A similar explanation is valid in the situation
investigated in \cite{STM2} and, in fact, the density plots of
Figure \ref{f4} are quite similar with those reported in
\cite{STM1}, \cite{STM2}. The number of interference fringes depends
on $e$ by the monotonicity of the exponent in Eq. (\ref{3.11}). The
fact that the density is higher along diagonals of the lattice,
which is striking in Figure \ref{f5}, is due to the fact that the
amplitude $\psi (e,\phi )$ has a minimum (maximum) at $\phi =\pi /4
({\rm resp\; }0)$; at $e=1.4$, the ratio of $|g_-|^2$ on the
diagonal and on the axis (at the same distance from the origin)
equals $\psi (1.4,0)/\psi (1.4,\pi /4)=2.264$; at $e=0.3$, this
ratio equals only $1.127$. This explains, at least for one contact
($t_1=0$), why the density is larger on the diagonals and why the
effect is better seen at the larger energy (Figures \ref{f3}c,
\ref{f4}c).

\bigskip
One expects a dependence of $\delta ^{(1)}(e;x)$ on the distance
$d_1$ between the two points of $S_1$, as the larger $d_1$, the less
correlated are the electrons incident at two distant contacts. The
calculations show that the visibility of the fringes decreases with
increasing $d_1$ (see Figure \ref{f5}, panel b). However, the
interference pattern is conserved even in the limit $d_1\to \infty$.
Indeed, in this limit $P(e)|_{S_1\times S_1}$ becomes diagonal,
hence, $m_{\rm tr}(e)=P(e)_{0,0}\lim{Q_+^{(2,1)}(e)}\cdot
\lim{Q_-^{(1,2)}(e)}$, which is by no means diagonal.

\begin{figure}[h]
\begin{center}
a\includegraphics[width=7cm]{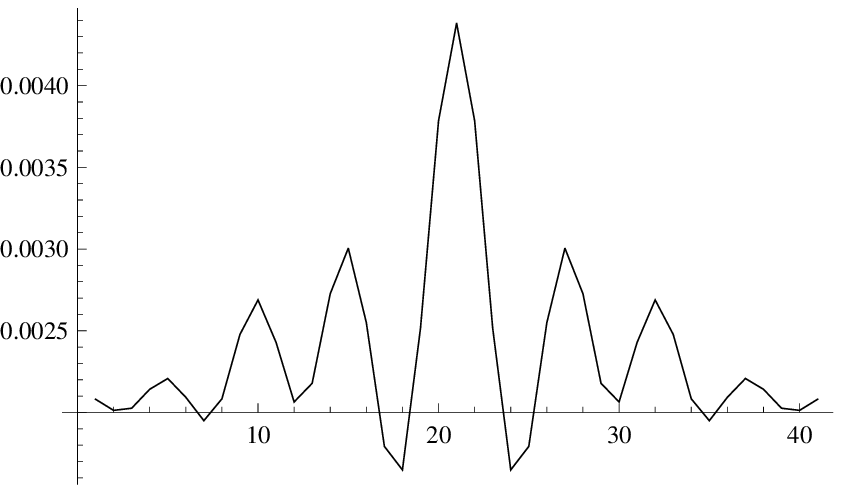}\
b\includegraphics[width=7cm]{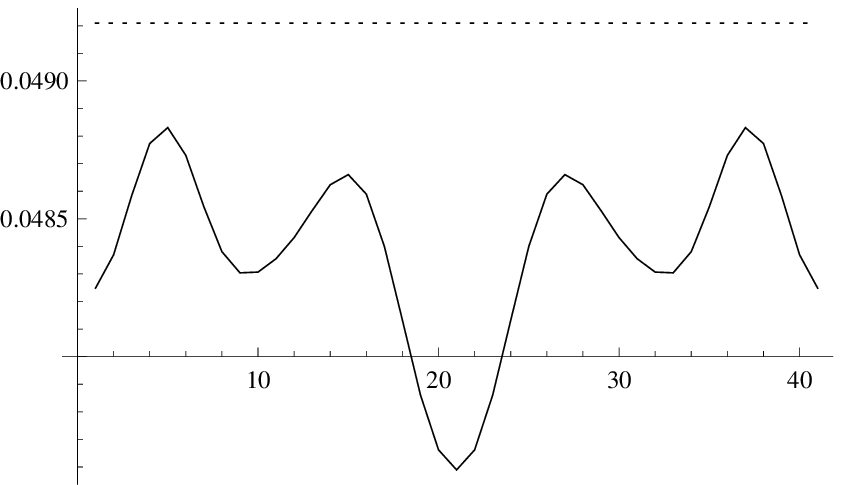}\\
\vskip 0.5cm c\includegraphics[width=7cm]{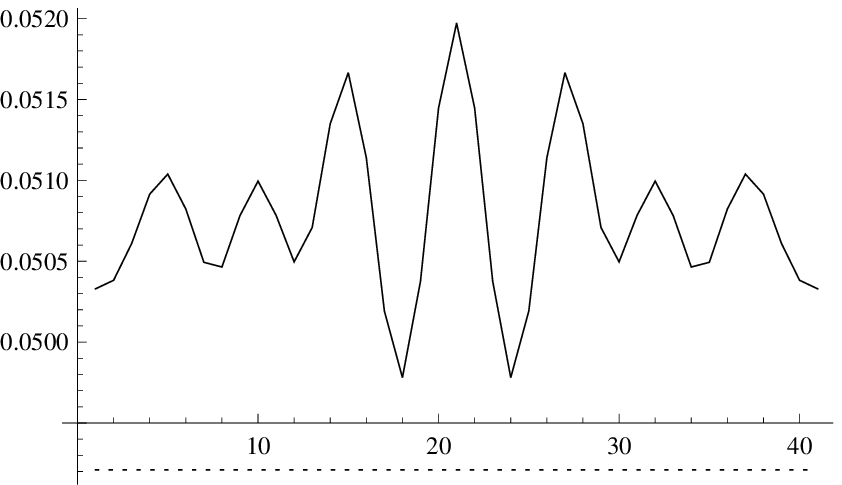}\\
\end{center}
\caption{Local densities on the line $x=(i,19), i=1,...,40$: a.
Transmitted electrons ($\mu _1=1.4$); b. Reflected electrons ($\mu
_2=0.3$); c. The (total) local density in the stationary state.
Dashed, the initial equilibrium density is represented.}\label{f6}
\end{figure}

The density distribution in $\mathbb{L}_2$ in the stationary state
is the sum $d(x)=\int _0^{1.4}\delta ^{(1)}(e;x){\rm d}e+\int
_0^{0.3}\delta ^{(2)}(e;x){\rm d}e$. As the position (and number) of
the fringes in $\delta ^{(i)}(e,x)$ depends on $e$, we expect the
interference pattern of $d(x)$ to have worse visibility. This is
confirmed by the computation of $d(x)$. We fixed, as before,
temperature to 0 and $\mu _1=1.4,\, \mu _2=0.3$. In a density plot
of $d(x)$, fringes are indeed hardly visible. Indeed, in the
intermediate region where fringes are more pronounced, $d(x)$ is
close to the initial equilibrium density of the second reservoir,
$\rho ^{(2)}_{\rm eq}=0.0492$ and has very small oscillations (of
order $10^{-3}$) around this value. In order to better exhibit the
existence of fringes, we plotted the two terms and their sum on the
line $x=(i,19), i=1,...,40$ in Figure \ref{f6}.

\subsection{The particle current}\label{s3.3}

\begin{figure}
\begin{center}
\includegraphics[width=8cm]{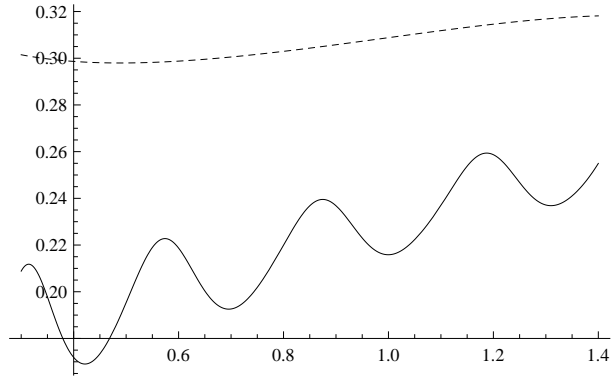}
\caption{Graph of $j(e)$}\label{f7}
\end{center}
\end{figure}

The total current flowing from reservoir 1 is given by Eq.
(\ref{2.24}). Under the assumptions of the previous subsection, in
particular $\beta_1=\beta_2=\infty$, $\mu _1=1.4,\, \mu _2=0.3$, it
is easy to see that
\begin{equation}\label{3.18}
J_{\mathbb{L}_1}=\int\limits_{0.3}^{1.4}j(e){\rm d}e;\; j(e)=2\pi
{\rm tr}_2\left[ {m_{\rm tr}(e)P(e)|_{S_2\times S_2}}\right] ,
\end{equation}
with $m_{\rm tr}(e)$ the same matrix as in Eq. (\ref{3.15}). We
report the calculation for the symmetric case $t_1=t_2=1$ and
$d_1=1$, $d_2=20$: $J_{\mathbb{L}_1}=0.2416$. It is interesting to
exhibit the energy resolution of the current, as well. The plot of
$j(e)$, represented in Figure \ref{f7}, shows oscillations,
indicating that tuning the energy may result in higher
conductivities. The dotted line in Figure \ref{f7} is the plot of
$2j_0(e)$, where $j_0(e)$ corresponds to one contact ($t_1=0,\;
t_2=1$, or viceversa). The latter plot, which is what Ohm's law
would predict, has no structure and definitely exceeds $j(e)$ over
the whole domain.

\begin{figure}
\begin{center}
a\includegraphics[width=7cm]{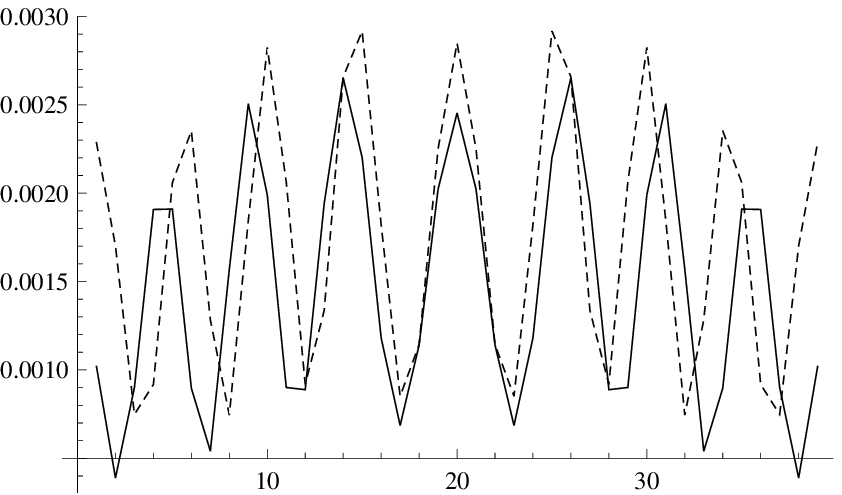}\
b\includegraphics[width=7cm]{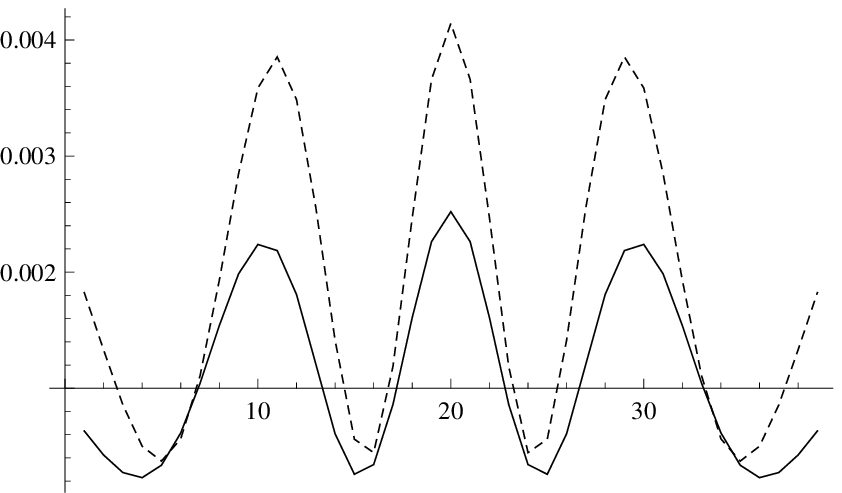}\\
\vskip 0.5cm c\includegraphics[width=7cm]{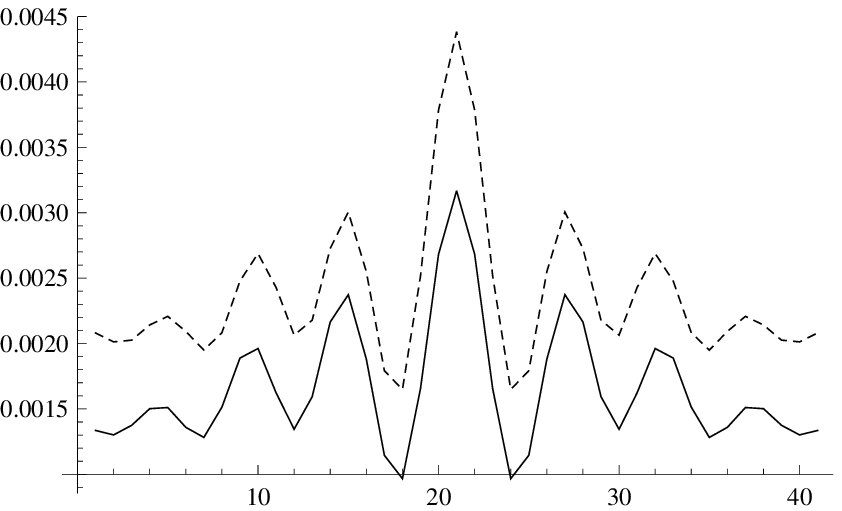} \caption{The
local "transmitted" current: a. at energy $e=1.4$; b. at energy
$e=0.3$; c. integrated on the energy range $(0,1.4)$, across the
vertical bonds $\{ (i,19),(i,20)\}$. Dashed, the local density in
points $(i,19)$ is plotted.}\label{f8}
\end{center}
\end{figure}

Interference effects are clearly seen in the distribution of the
local currents in a neighborhood of the contacts. We computed, using
Eq. (\ref{2.20}), the local currents in the stationary state in the
second reservoir in the same $40\times 40$ square around $S_2$, for
the same setting and parameter values as before. For $\{ x,y\}$ a
bond of nearest neighbors in $\mathbb{L}_2$, one has, after
integration over the energy shells,
$$j_{x,y}=\int\limits_0^{0.3}j_{x,y}^{({\rm ref})}(e){\rm
d}e+\int\limits_0^{1.4}j_{x,y}^{({\rm tr})}(e){\rm d}e,$$ where:
\begin{equation}\label{3.19}
\begin{array}{l}
j_{x,y}^{({\rm tr})}(e)=-\Im{\left(
V_-(e,x),m_{\rm tr}(e) V_-(e,y)\right) _{\mathbb{C}^2}}, \\
j_{x,y}^{({\rm ref})}(e)=2\Im{\left(
V_-(e,x),Q_+^{(2,2)}(e)V^{(0)}(e,y)\right)
_{\mathbb{C}^2}}-\Im{\left( V_-(e,x),m_{\rm ref}(e) V_-(e,y)\right)
_{\mathbb{C}^2}}.
\end{array}
\end{equation}

Figure \ref{f8} is a plot of the local currents of transmitted
electrons along nearest-neighbor bonds crossing a horizontal line in
the region of fringes: a. of fixed energy $e= 1.4$, i.e.
$j_{x,y}^{({\rm tr})}(1.4)$; b. of fixed energy $e=0.3$,
$j_{x,y}^{({\rm tr})}(0.3)$; c. integrated over the whole energy
range, i.e. $j_{x,y}^{({\rm tr})}=\int\limits_0^{1.4}j_{x,y}^{({\rm
tr})}(e){\rm d}e$. We took $\{ x,y\} = \{ (i,19),(i,20)\}  ,\;
i=1,...,40$. We represented also the local density on that line, in
order to demonstrate the correlation of the two quantities: currents
are larger on bonds starting from high-density sites. The
"reflected" local currents $j_{x,y}^{({\rm ref})}$ across the same
line show the same kind of correlation with the local density, the
major difference being that the currents are negative, i.e.
electrons enter the neighborhood of $S_2$.

\begin{figure}
\begin{center}
\includegraphics[width=8cm]{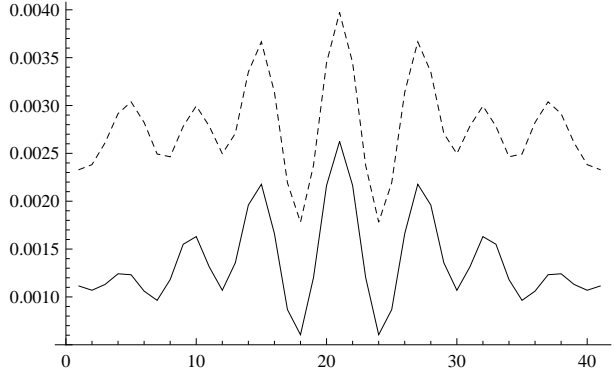}
\caption{The local currents in the stationary state across the line
of bonds $\{ (i,19),(i,20)\}  ,\; i=1,...,40$. Dotted is the local
density (shifted downwards with 0.048) at sites $(i,19)$.}\label{f9}
\end{center}
\end{figure}

Finally, we plotted in Figure \ref{f9} the total (transmitted +
reflected) local currents in the stationary state, $j_{x,y}$, on the
same line of bonds $\{ x,y\} =\{ (i,19),(i,20)\}  ,\; i=1,...,40$,
along with a copy of the plot of the local density at $x$ (Figure
\ref{f6}c). As the variation of $j_{x,y}^{({\rm tr})}$ is an order
of magnitude larger than that of $j_{x,y}^{({\rm ref})}$, the
picture is quite similar to that of $j_{x,y}^{({\rm tr})}$ (Figure
\ref{f8}c), and the same correlation with the total local density is
observed.
\section{Conclusion}\label{s4}
\setcounter{equation}{0}

We have performed a detailed study of the stationary state, for the
model under consideration, by calculating the expectations of
various local observables, namely, the number of particles in a
lattice site and the particle current along a nearest-neighbor bond
(in a particular geometric arrangement and tunneling constants).
These expectations exhibit a peculiar dependence on the position of
the lattice site/bond relative to the two contacts, putting into
evidence interference patterns consisting of fringes of high local
density, respectively local current. We chose to calculate this
space dependence in the less populated of the two systems. Both the
density and current profiles are sums of contributions from
transmitted and reflected particles (corresponding to one of the
systems being unpopulated). The two contributions yield
qualitatively different patterns: the fringes of the transmitted
particles are similar to hyperbolae with foci in the contacts and
they disappear as one of the contacts is suppressed, while the
fringes of the reflected particles are similar to circles centered
in the contacts and fringes are present even in the case of one
contact. We explained this peculiarity, as well as the dependence of
the pattern on the various parameters. The space dependence of the
sum of the two contributions turns out, in our case, to be dominated
by the transmitted particles: hyperbolic fringes of higher density
along which particles flow from the contacts to infinity.

New interference effects are to be expected when having, instead of
direct tunneling, some finite intermediate quantum system, possibly
with a Coulomb repulsion included, in which case the scattering has
a resonant structure. For small tunneling constants, resonances come
close to the energy levels of the intermediate quantum system, with
the effect that only certain energy channels are open. In this case,
the fixed-energy calculations performed above become relevant. We
propose to follow the subject in another publication.

\bigskip
\textbf{\large Acknowledgments}

The authors acknowledge financial support from Romanian National
Authority for Scientific Research via the programs "Nucleu",
contracts NIFIN 3 PN 09 37 and LAPLAS 3 PN 09 39.


\newpage
\begin{thebibliography}{99}

\bibitem{AJPP} W. Aschbacher, V. Jaksic, Y. Pautrat, and C.-A. Pillet 2007 {\it
J.Math.Phys.} {\bf 48}  032101

\bibitem{abb} Angelescu N, Bundaru M and Bundaru R 2008  {\it Quasi-free Quantum Statistical Models for Tunneling Junctions}, in {\it Topics in Applied Mathematics and
Mathematical Physics} (Bucure\c sti: Ed. Acad. Rom\^{a}ne ) pp~11-44

\bibitem{BR} Bratteli O and Robinson D W 1979  {\it Operator Algebras and Quantum Statistical Mechanics I} (New York: Springer )

\bibitem{STM1} Crommie M F, Lutz C P and Eigler D M 1993
{\it  Nature} (London) {\bf 363}  524

\bibitem{STM2} Heller E J, Crommie M F, Lutz C P and Eigler D M 1994
{\it Nature} (London) {\bf 369} p~464

\bibitem{nano} Sun G F, Liu Y, Qi Y, Jia J F, Weinert M and Li L 2010
{\it Nanotechnology} {\bf 21} 435401.

\bibitem{focus} Sentef M, Kampf A P, Hembacher S and Mannhart J 2006
{\it Phys.Rev. B} {\bf 74}  153407.

\bibitem{BE} H. Bateman (ed.)  A. Erdélyi (ed.) 1953  \textit{Higher transcendental functions ,
1. The gamma function. The hypergeometric functions. Legendre
functions} (New York: McGraw-Hill)

\bibitem{Erd} Erd\'{e}lyi A 1956 \textit{Asymptotic Expansions}
( New York: DoverPubl. Inc.)

\bibitem{v} Vainberg B R 1989 \textit{Asymptotic Methods in Equations of
Mathematical Physics} (New York: Gordon and Breach )

\end{thebibliography}
\end{document}